\documentclass[11pt, onecolumn, twoside]{IEEEtran}

\usepackage{graphicx}
\usepackage{amsmath,amssymb}
\usepackage{cite}

\DeclareMathOperator*{\argmin}{arg\,min}

\newtheorem{theorem}{Theorem}
\newtheorem{proposition}{Proposition}

\begin{document}

\title{Reliable Crowdsourcing for Multi-Class Labeling using Coding Theory}

\author{Aditya Vempaty, Lav~R.~Varshney, and Pramod~K.~Varshney%
\thanks{A.~Vempaty and P.~K.~Varshney are with the Department of Electrical Engineering and Computer Science, Syracuse University, Syracuse, NY, 13244 USA (e-mail: \{avempaty,varshney\}@syr.edu).}
\thanks{L.~R.~Varshney is with the Department of Electrical and Computer Engineering and with the Coordinated Science Laboratory, University of Illinois at Urbana-Champaign, Urbana, IL 61801, USA (e-mail: varshney@illinois.edu). He was previously with the IBM Thomas J.~Watson Research Center, Yorktown Heights, NY, USA.}
\thanks{Portions of the material in this paper were first presented in \cite{VempatyVV2013}. This work was supported in part by the Center for Advanced Systems and Engineering (CASE) at Syracuse University, ARO under Grant W911NF-09-1- 0244, AFOSR under Grants FA9550-10-1-0458 and FA9550-10-1-0263.}}

\maketitle

\begin{abstract}
Crowdsourcing systems often have crowd workers that perform unreliable work on the task they are assigned. In this paper, we propose the use of error-control codes and decoding algorithms to design crowdsourcing systems for reliable classification despite unreliable crowd workers. Coding-theory based techniques also allow us to pose easy-to-answer binary questions to the crowd workers.  We consider three different crowdsourcing models: systems with independent crowd workers, systems with peer-dependent reward schemes, and systems where workers have common sources of information. For each of these models, we analyze classification performance with the proposed coding-based scheme. We develop an ordering principle for the quality of crowds and describe how system performance changes with the quality of the crowd. We also show that pairing among workers and diversification of the questions help in improving system performance. We demonstrate the effectiveness of the proposed coding-based scheme using both simulated data and real datasets from Amazon Mechanical Turk, a crowdsourcing microtask platform.  Results suggest that use of good codes may improve the performance of the crowdsourcing task over typical majority-voting approaches.
\end{abstract}


\section{Introduction}
Conventional studies of human decision making involve decisions made by individuals such as military commanders or company presidents, or by a small number of individuals such as committees or cabinets operating under decision fusion rules such as majority voting.  Behavioral scientists have studied human decision making from a cognitive viewpoint whereas economists have studied it from a decision theory viewpoint.  When agents are rational and knowledgeable about the problem domain, team decision theory provides well-developed insights \cite{MarschakR1972}.  With the emergence of widespread and inexpensive computing and networking infrastructure, new paradigms for human participation, such as crowdsourcing, have arisen for distributed inference tasks \cite{Bollier2011,TapscottW2006,TapscottW2010} and are becoming very prevalent in the specialized and globalized knowledge economy. 

Although both crowdsourcing and conventional team decision making \cite{MarschakR1972} involve human decision makers, there are three major differences. First, the number of participants involved in crowdsourcing is usually large.  Second, contrary to traditional mathematical models of team decision making \cite{Varshney1997}, members of the crowd may be unreliable or malicious \cite{IpeirotisPW2010,Lease2011,Varshney2012d} especially since they are often anonymous \cite{RossISZT2010,DownsHSC2010}.  Third, workers may not have sufficient domain expertise to perform
full classification and may only be able to make simpler binary distinctions \cite{BransonWSBWPB2010}. These differences give rise to a multitude of new system design challenges when crowdsourcing is employed for tasks with varying quality requirements and time deadlines.  The main challenge, however, is \emph{quality control} to ensure reliable crowd work \cite{IpeirotisPW2010} with provable performance guarantees. 

In typical networked sensing systems, heterogeneous sensors are used to observe a phenomenon and to collaboratively make inferences (detection, classification, or estimation), where algorithms at the sensors or at a fusion center are derived to operate optimally or near-optimally. In large sensor networks consisting of inexpensive battery-powered sensors with limited capabilities, one key issue has been to maintain inference quality in the presence of faulty nodes or communication errors.  Recently an innovative marriage of concepts from coding theory and distributed inference has been proposed \cite{WangHVC2005,YaoCWHV2007}, where the goal is to jointly maximize classification performance and system fault tolerance by jointly designing codes and decision rules at the sensors.  In the present work, we apply distributed inference codes \cite{WangHVC2005} to crowdsourcing tasks like classification.  This is consistent with popular uses of crowdsourcing microtask platforms such as Amazon Mechanical Turk.

Workers often find microtasks tedious and due to lack of motivation fail to generate high-quality work \cite{QuinnB2011}.  It is, therefore, important to design crowdsourcing systems with sufficient incentives for workers \cite{LawV2011}.  The most common incentive for workers is monetary reward, but in \cite{Varshney2012e}, it has been found that intrinsic factors such as the challenge associated with the task was a stronger motivation for crowd workers than extrinsic factors such as rewards. Unfortunately, it has been reported that increasing financial incentives increases the number of tasks which the workers take part in but not the per task quality \cite{MasonW2009}.  Recent research, however, suggests that making workers' rewards co-dependent on each other can significantly increase the quality of their work. This suggests the potency of a \emph{peer-dependent reward scheme} for quality control \cite{HuangF2013}. In a \emph{teamwork-based scheme}, paired workers are rewarded based on their average work whereas in a \emph{competition-based} scheme, the paired worker who performs the best gets the entire reward. Herein, we develop a mathematical framework to evaluate the effect of such pairings among crowd workers. 

Another interesting phenomenon in crowdsourcing is \emph{dependence} of observations among crowd workers \cite{QiAHH2013}.  Crowds may share common sources of information, leading to dependent observations among the crowd workers performing the task. Common sources of information have oft-been implicated in the publishing and spread of false information across the internet, e.g. the premature Steve Jobs obituary, the second bankruptcy of United Airlines, and the creation of black holes by operating the Large Hadron Collider \cite{Berti-EquilleSDMS2009}. A graphical model may be used to characterize such dependence among crowd workers \cite{QiAHH2013}; herein we also consider coding-based classification under dependent observations. It should be noted that only independent observations were considered in the original work on coding-based classification \cite{WangHVC2005,YaoCWHV2007}.

Since quality control is a central concern for crowdsourcing \cite{IpeirotisPW2010}, previous work considered numerical analysis methods \cite{Grier2011b} and binary classification tasks \cite{KargerOS2011a,KargerOS2011b}. In \cite{KargerOS2011a,KargerOS2011b}, the authors considered the problem of task allocation in a crowdsourcing system; an iterative algorithm based on belief propagation was proposed for inferring the final answer from the workers' responses. This algorithm was shown to perform as well as the best possible algorithm. They also provided numerical results for large system size and showed that their approach outperforms the majority-based approach.  Extensions to a multi-class labeling task \cite{KargerOS2013} provided an algorithm to obtain the best tradeoff between reliability and redundancy.  This algorithm was based on low-rank approximation of weighted adjacency matrices for random regular bipartite graphs used for task allocation. 

We focus on crowdsourcing for $M$-ary classification tasks, such as multi-class object recognition from images into fine-grained categories \cite{BransonWSBWPB2010}.  We aim to design the system to minimize misclassification probability by using distributed classification codes and a minimum Hamming distance decoder such that workers need only answer binary questions.  We demonstrate the efficacy of this coding-based approach using simulations and through real data from Amazon Mechanical Turk \cite{SnowOJN2008}, a paid crowdsourcing microtask platform.  We analyze the approach under different crowdsourcing models including the peer-dependent reward scheme \cite{HuangF2013} and the dependent observations model \cite{QiAHH2013}. In the process, an ordering principle for the quality of crowds is also developed. For systems with peer-dependent reward schemes, we observe that higher correlation among workers results in performance degradation. Further, if the workers also share dependent observations due to common sources of information, we show that the system performance deteriorates as expected. However, we also observe that when the observations become independent, the performance gain due to our coding-based approach over the majority-vote approach increases. 

Sec.~\ref{sec:setup} develops a mathematical model of the crowdsourcing problem and proposes the coding-based approach. Sec.~\ref{sec:iid} considers a crowdsourcing system with independent workers.  The peer-dependent reward scheme is introduced in Sec.~\ref{sec:peer} and the system is further generalized in Sec.~\ref{sec:dep} by allowing dependence among crowd worker observations. For each of these models, in turn, misclassification performance expressions for both coding- and majority-based approaches are derived. Examples demonstrate that systems with good codes outperform systems that use majority voting.  Experimental results using real data from Amazon Mechanical Turk are also provided wherever applicable. Concluding remarks are provided in Sec.~\ref{sec:conc}.

\section{Coding for Crowdsourcing}
\label{sec:setup}	
In this section, we discuss the basic concept of using error-correcting codes to achieve reliable classification in a crowdsourcing system. We first describe the Distributed Classification Fusion using Error-Correcting Codes (DCFECC) approach for Wireless Sensor Networks (WSNs) proposed by Wang et. al in \cite{WangHVC2005} which serves as the mathematical basis for the idea proposed in this paper. 

\subsection{Distributed Classification Fusion using Error-Correcting Codes}
\label{sec:DCFECC}

\begin{figure}
  \centering
  \includegraphics[width = 3.5in, height=!]{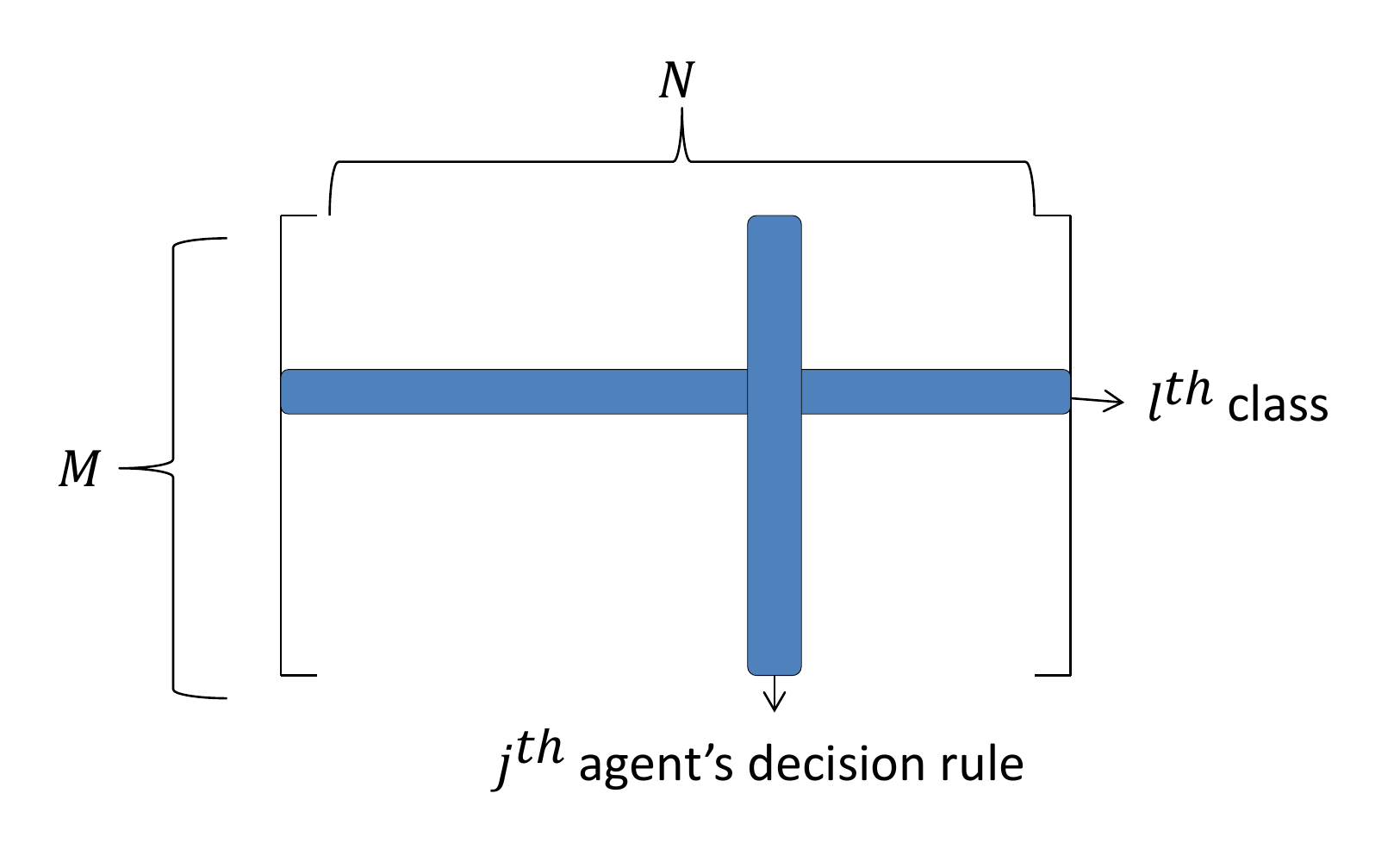}
  \caption{Code matrix representation of the DCFECC approach.}
  \label{fig:codematrix}
\end{figure}

DCFECC hinges on the simple idea of representing a distributed classification problem using a binary code matrix $\mathbf{A}$. If there are $M$ hypotheses (classes) $H_0, H_1,\ldots, H_{M-1}$ that need to be distinguished by $N$ agents, the code matrix $\mathbf{A}$ is of size $M \times N$. Each row, a codeword of $\mathbf{A}$, corresponds to one of the possible hypotheses and the columns represent the decision rules of the agent as depicted in Fig.~\ref{fig:codematrix}. 

Given this code matrix, agent $j$ sends its binary decision ($u_j \in \{0,1\}$). The fusion center receives the $N$-bit vector, $\mathbf{u}=[u_1, \cdots, u_N]$ and makes the final classification decision by using minimum Hamming distance based fusion. In other words, the fusion rule used by the fusion center is to decide $H_i$, where 
\begin{equation}
i = \argmin_{0 \leq l \leq M-1} d_H(\mathbf{u}, \mathbf{r}_l),
\end{equation} 
$d_H(\mathbf{x},\mathbf{y})$ is the Hamming distance between $\mathbf{x}$ and $\mathbf{y}$, and $\mathbf{r}_l=(a_{l1}, \cdots, a_{lN})$ is the row of $\mathbf{A}$ corresponding to hypothesis $H_l$. The tie-break rule is to randomly pick a codeword from those with the same smallest Hamming distance to the received vector. Due to the minimum Hamming-distance based fusion scheme, the DCFECC approach can also handle missing data. When an agent does not return any decision, its contribution to the Hamming distance between the received vector and every row of the code matrix is the same and, therefore, the row corresponding to the minimum Hamming distance remains unchanged. The error-correction property of the code matrix $\mathbf{A}$ provides fault-tolerance capability, as shown in \cite{WangHVC2005}. 

As is evident from the above discussion, classification performance depends on the code matrix $\mathbf{A}$ since it is used for designing the local decision rules as well as the final classification fusion by the fusion center. This code matrix is designed to minimize the misclassification probability. Two heuristic methods have been typically used for code design: cyclic column replacement and simulated annealing. In cyclic column replacement, we start with an initial matrix and replace the matrix column wise to minimize the misclassification probability. This approach has lower computational complexity but might result in a local optimum. The simulated annealing approach searches for a globally optimal solution at the expense of high computational complexity. The exact expression, which characterizes the performance, and the optimal code matrix depend on the application considered. 

\subsection{Reliable Classification using Crowds}
\label{sec:rel_crowds}
We now describe how the DCFECC approach can be used in crowdsourcing systems to design the questions to be posed to the crowd workers. As an example, consider an image to be classified into one of $M$ fine-grained categories.  Since object classification is often difficult for machine vision algorithms, human workers may be used for this task.  In a typical crowdsourcing microtask platform, a task manager creates simple tasks for the workers to complete, and the results are combined to produce the final result.  Due to the low pay of workers and the difficulty of tasks, individual results may be unreliable. Furthermore, workers may not be qualified to make fine-grained $M$-ary distinctions, but rather can only answer easier questions.  Therefore, in our approach, codes are used to design microtasks and decoding is performed to aggregate responses reliably. 

Consider the task of classifying a dog image into one of four breeds: Pekingese, Mastiff, Maltese, or Saluki.  Since workers may not be canine experts, they may not be able to directly classify and so we should ask simpler questions.  For example, the binary question of whether a dog has a snub nose or a long nose differentiates between \{Pekingese, Mastiff\} and \{Maltese, Saluki\}, whereas the binary question of whether the dog is small or large differentiates between \{Pekingese, Maltese\} and \{Mastiff, Saluki\}.  Using a code matrix, we now show how to design binary questions for crowd workers that allow the task manager to reliably infer correct classification even with unreliable workers. 
 
As part of modeling, let us assume that worker $j$ decides the true class (local decision $y_j$) with probability $p_j$ and makes the wrong local classification with uniform probability:
\begin{equation}
\label{eq:obs_model}
p(y_j|H_m)=
\begin{cases}
p_j & \mbox{if } y_j=m \\
\frac{1-p_j}{M-1} & \mbox{otherwise,}
\end{cases}
\end{equation}

\begin{figure}
  \centering
  \includegraphics[width = 4.25in, height=!]{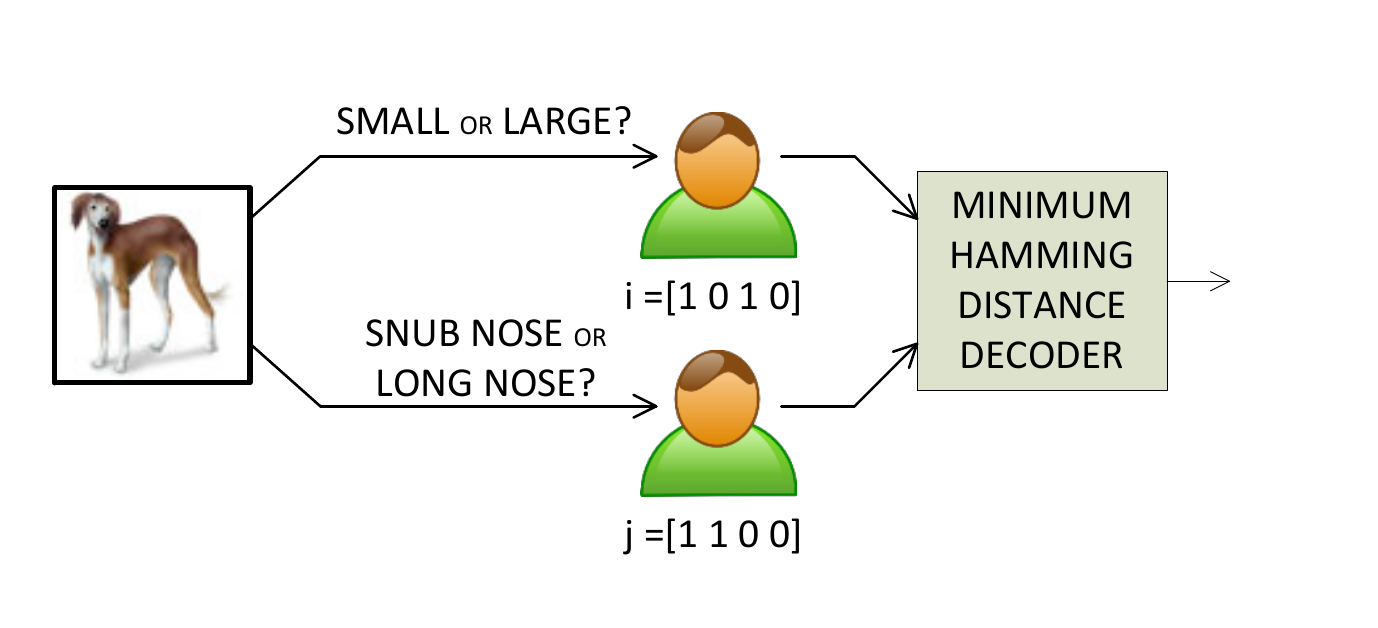}
  \caption{A schematic diagram showing binary questions posed to workers and the decoding rule used by the task manager.}
  \label{fig:diagram}
\end{figure}

Note that, the uniform noise model of this paper can be regarded as the ``worst case" in information. In such a case, this would relate to an upper bound on the performance of the system.
For every worker $j$, let $a_j$ be the corresponding column of $\mathbf{A}$ and recall hypothesis $H_l \in \{H_0, H_1,\cdots, H_{M-1}\}$ is associated with row $l$ in $\mathbf{A}$.  The local workers send a binary answer $u_j$ based on decision $y_j$ and column $a_j$.  An illustrative example is shown in Fig.~\ref{fig:diagram} for the dog breed classification task above.  Let the columns corresponding to the $i$th and $j$th workers be $a_i=[1 0 1 0]'$ and $a_j=[1 1 0 0]'$ respectively. The $i$th worker is asked: ``Is the dog small or large?'' since she is to differentiate between the first (Pekingese) or third (Maltese) breed and the others.  The $j$th worker is asked: ``Does the dog have a snub nose or a long nose?'' since she is to differentiate between the first two breeds (Pekingese, Mastiff) and the others. These questions can be designed using taxonomy and dichotomous keys \cite{RockerYYPZ2007}. Knowing that Pekingese and Maltese are small dogs while the other two breeds are large, we can design the appropriate question as ``Is the dog small or large?'' for $i$th worker whose corresponding column is $a_i=[1 0 1 0]'$. The task manager makes the final classification as the hypothesis corresponding to the codeword (row) that is closest in Hamming distance to the received vector of decisions.

\subsection{Unreliable Workers}
Although distributed classification in sensor networks and in crowdsourcing are structurally similar, an important difference is the anonymity of crowds. Since crowd workers are anonymous, we cannot identify the specific reliability of 
any specific worker as could be done for a sensor.  Hence, we assume that each worker $j$ in the crowd has an associated reliability $p_j$, drawn from a common distribution that characterizes the crowd.  Herein, we consider three different crowd models that generate crowd reliabilities: individual and independent crowd workers; crowd workers governed by peer-dependent reward schemes; and crowd workers with common sources of information.  In the following sections, we analyze each of these models to evaluate the proposed coding-based scheme.

\section{Crowdsourcing System with Individual Crowd Workers}
\label{sec:iid}
In this section, we analyze the basic crowdsourcing system where independent crowd workers perform the task individually and are rewarded based on their decision only.

\subsection{Model}
The system consisting of individual and independent workers can be modeled as one where the workers' reliabilities are drawn i.i.d.\ from a specific distribution. Two crowd reliability models namely a spammer-hammer model and a beta model are considered herein. In a spammer-hammer model, the crowd consists of two kinds of workers: spammers and hammers. Spammers are unreliable workers that make a decision at random whereas hammers are reliable workers that make a decision with high reliability. The quality  of the crowd, $Q$, is governed by the fraction of hammers. In a beta model, the reliabilities of workers are drawn from a beta distribution with parameters $\alpha$ and $\beta$. 

\subsection{Performance Characterization}
\label{sec:perf}
Having defined a coding-based approach to reliable crowdsourcing, we determine its performance in terms of average misclassification probability for classification under minimum Hamming distance decoding.  Suppose $N$ workers take part in an $M$-ary classification task. Let $\underline{p}$ denote the reliabilities of these workers, such that $p_j$ for $j = 1,\ldots,N$ are i.i.d.\ random variables with mean $\mu$.  We define this to be an $(N,M,\mu)$ crowdsourcing system.

\begin{proposition}
\label{prop1}
Consider an $(N,M,\mu)$ crowdsourcing system. The expected misclassification probability using code matrix $\mathbf{A}$ is:
\begin{equation}
\label{eq:err_coding}
P_e(\mu)=\frac{1}{M}\sum_{\underline{i},l}\prod_{j=1}^N \Bigg[\left(\mu a_{lj} + \frac{(1-\mu)}{(M-1)}\sum_{k\neq l}a_{kj}\right)(2i_j-1) +(1-i_j)\Bigg]C_{\underline{i}}^l \mbox{,}
\end{equation}
where $\underline{i}=[i_1, \cdots, i_N] \in \{0, 1\}^N$ is the received codeword and $C^l_{\underline{i}}$ is the cost associated with a global decision $H_l$ when the received vector is $\underline{i}$. This cost is:
\begin{equation}
\label{eq:cost}
C_{\underline{i}}^l=
\begin{cases}
1-\frac{1}{\varrho} &\mbox{if } \underline{i} \mbox{ is in decision region of } H_l\\
1 & \mbox{otherwise.}
\end{cases} 
\end{equation}
where $\varrho$ is the number of decision regions\footnote{For each $H_l$, the set of $\underline{i}$ for which the decision $H_l$ is taken is called the decision region of $H_l$.} $\underline{i}$ belongs to; $\varrho$ can be greater than one when there is a tie at the task-manager and the tie-breaking rule is to choose one of them randomly.
\end{proposition}
\begin{IEEEproof}
Let $P_{e,\underline{p}}$ denote the misclassification probability given the reliabilities of the $N$ workers. Then, if $u_j$ denotes the bit sent by the worker $j$ and the global decision is made using the Hamming distance criterion:
\begin{equation}
P_{e,\underline{p}}=\frac{1}{M}\sum_{\underline{i},l}P(\underline{u}=\underline{i}|H_l)C_{\underline{i}}^l\mbox{.}
\end{equation}
Since local decisions are conditionally independent, $P(\underline{u}=\underline{i}|H_l)=\prod_{j=1}^N P(u_j=i_j|H_l)$.
Further, 
\begin{align*}
P(u_j=i_j|H_l) &= i_jP(u_j=1|H_l)+(1-i_j)P(u_j=0|H_l)\\
&=(1-i_j)+(2i_j-1)P(u_j=1|H_l)\\
&=(1-i_j)+(2i_j-1)\sum_{k=1}^Ma_{kj}P(y_j=k|H_l)\\
&=(1-i_j)+\left(p_j a_{lj} + \frac{(1-p_j)}{(M-1)}\sum_{k\neq l}a_{kj}\right)(2i_j-1)
\end{align*}
where $y_j$ is the local decision made by worker $j$. Since reliabilities $p_j$ are i.i.d.\ with mean $\mu$, the desired result follows.
\end{IEEEproof}

\subsubsection*{Performance Bound}
Yao et al.\ provide performance analysis for the distributed $M$-ary classification fusion system with minimum Hamming distance fusion \cite{YaoCWHV2007}. We state their result here without proof. This result can be used in the context of distributed $M$-ary classification using $N$ workers with reliabilities $\{p_j\}_{j=1}^N$ and a code matrix $\mathbf{A}$ for coding-based classification.
 
\begin{proposition}[{\cite{YaoCWHV2007}}] Let $P_e$ be the probability of minimum Hamming distance fusion misclassification error given as
\begin{equation}
P_e \triangleq \frac{1}{M}\sum_{i=1}^{M-1}P(\text{fusion decision} \neq H_i|H_i).
\end{equation}

If for every $l \neq i$
\begin{equation}
\label{eq:cond}
\sum_{\{j\in [1,\cdots,N]:a_{lj}\neq a_{ij}\}}E[z_{i,j}] =\sum_{j=1}^{N}(a_{lj}\oplus a_{ij})(2q_{i,j}-1) <0,
\end{equation}
where $0\leq l,i \leq M-1$,  $z_{i,j}  \triangleq 2(u_j \oplus a_{ij})-1$, $\oplus$ represents the `xor' operation and $q_{i,j}\triangleq P\{z_{i,j}=1|H_i\}$, then
\begin{equation}
P_e \leq \frac{1}{M}\sum_{i=0}^{M-1}\sum_{0 \leq l \leq M-1, l \neq i}\inf_{\theta \geq 0}\exp\left\{\sum_{j=1}{N}\log{(q_{i,j}e^\theta + (1-q_{i,j})e^{-\theta})^{a_{lj}\oplus a_{ij}}}\right\}.
\end{equation}
\end{proposition}

The proof of the proposition follows from large deviations theory \cite{YaoCWHV2007}. In crowdsourcing, the probabilities $q_{i,j}=P\{u_j\neq a_{ij}\}$ can be easily computed as:
\begin{eqnarray}
q_{i,j}&=&\sum_{l=0}^{M-1}(a_{ij}\oplus a_{lj})h_{l|i}^{(j)},
\end{eqnarray}
where $h_{l|i}^{(j)}$ is the probability that worker $j$ decides $H_l$ when the true hypothesis is $H_i$ and are given by
\begin{equation}
h_{l|i}^{(j)}=
\begin{cases}
p_j,	&	\text{$i=l$}\\
\frac{1-p_j}{M-1},	&	i\neq l \mbox{.}
\end{cases}
\end{equation}

\subsection{Majority Voting}
\label{sec:majority}
A traditional approach in crowdsourcing has been to use a majority vote to combine local decisions; we also derive its performance for purposes of comparison.  For $M$-ary classification, each worker's local decision is modeled as $\log_2 M $-bit valued, but since workers only answer binary questions, the $N$ workers are split into $\log_2 M $ groups with each group sending information regarding a single bit. For example, consider the dog breed classification task of Sec. \ref{sec:rel_crowds} which has $M=4$ classes. Let us represent the classes by 2-bit numbers as follows: Pekingese is represented as `00', Mastiff as `01', Maltese as `10', and Saluki as `11'. The $N$ crowd workers are split into 2 groups. In traditional majority vote, since the workers are asked $M$-ary questions, each worker first identifies his/her answer. After identifying his/her class, the first group members send the first bit corresponding to their decisions, while the second group members send the second bit of their decisions. The task manager uses a majority rule to decide each of the $\log_2{M}$ bits separately and concatenates to make the final classification. Suppose $N$ is divisible by $\log_2M$.
\begin{proposition}
\label{prop:maj_er_ind}
Consider an $(N,M,\mu)$ crowdsourcing system. The expected misclassification probability using majority rule is: 
\begin{equation}
\label{eq:err_majority}
P_e(\mu)=1-\frac{1}{M}\left[1+S_{\tilde{N},(1-q)}\left(\frac{\tilde{N}}{2}\right)-S_{\tilde{N},q}\left(\frac{\tilde{N}}{2}\right)\right]^{\log_2M},
\end{equation}
where $\tilde{N}=\frac{N}{\log_2M}$, $q=\frac{M(1-\mu)}{2(M-1)}$, and $S_{N,p}(\cdot)$ is the survival function (complementary cumulative distribution function) of the binomial random variable $\mathcal{B}(N,p)$.
\end{proposition}
\begin{IEEEproof}
In a majority-based approach, $\tilde{N}=\tfrac{N}{\log_2M}$ workers send information regarding the $i$th bit of their local
decision, $i=1,\ldots,\log_2 M$. For a correct global decision, all bits have to be correct. Consider the $i$th bit and let 
$P^i_{c,\underline{p}}$ be the probability of the $i$th bit being correct given the reliabilities of the $\tilde{N}$ workers 
sending this bit. Then,
\begin{equation}
\label{eq:Pc}
P^i_{c,\underline{p}}=\frac{P_d +1 - P_f}{2} \mbox{,}
\end{equation}
where $P_d$ is the probability of detecting the $i$th bit as `1' when the true bit is `1' and $P_f$ is the probability of detecting the
$i$th bit as `1' when the true bit is `0'. Note that `0' and `1' are equiprobable since all hypotheses are equiprobable. Under
majority rule for this $i$th bit,
\[
P_d=\sum_{j=\lfloor\frac{\tilde{N}}{2}+1\rfloor}^{\tilde{N}}\sum_{\forall G_j}\prod_{k\in G_j}\left(1-\frac{M(1-p_k)}{2(M-1)}\right)\prod_{k \notin G_j} \frac{M(1-p_k)}{2(M-1)} \mbox{,}
\]
where $G_j$ is a set of $j$ out of $\tilde{N}$ workers who send bit value `1' and $\frac{M(1-p_k)}{2(M-1)}$ is the probability of the $k$th worker making a wrong decision for the $i$th bit. Similarly,
\[
P_f=\sum_{j=\lfloor\frac{\tilde{N}}{2}+1\rfloor}^{\tilde{N}}\sum_{\forall G_j}\prod_{k\in G_j}\frac{M(1-p_k)}{2(M-1)}\prod_{k \notin G_j}\left(1-\frac{M(1-p_k)}{2(M-1)}\right) \mbox{.}
\]
Now the overall probability of correct decision is given by $P_{c,\underline{p}}=\prod_{i=1}^{\log_2M}P_{c,\underline{p}}^i$. Since reliabilities are i.i.d., the expected probability of correct decision $P_c$ is:
\begin{equation}
\label{eq:prod}
P_c = \prod_{i=1}^{\log_2M}E[P^i_{c,\underline{p}}],
\end{equation} 
where expectation is with respect to $\underline{p}$. Since reliabilities are i.i.d.:
\begin{small}
\begin{eqnarray}
\label{eq:Pd_expected}
E[P_d]=\sum_{j=\lfloor\frac{\tilde{N}}{2}+1\rfloor}^{\tilde{N}} \binom{\tilde{N}}{j} (1-q)^jq^{(\tilde{N}-j)}=S_{\tilde{N},(1-q)}\left(\frac{\tilde{N}}{2}\right)\mbox{,}\\
E[P_f]=\sum_{j=\lfloor\frac{\tilde{N}}{2}+1\rfloor}^{\tilde{N}} \binom{\tilde{N}}{j} q^j(1-q)^{(\tilde{N}-j)}=S_{\tilde{N},q}\left(\frac{\tilde{N}}{2}\right) \mbox{.}
\label{eq:Pf_expected}
\end{eqnarray}
\end{small}
Using \eqref{eq:Pc}, \eqref{eq:prod}, \eqref{eq:Pd_expected}, and \eqref{eq:Pf_expected}, we get the desired result.
\end{IEEEproof}

\subsection{Performance Evaluation}
\label{sec:res}

The performance expressions derived in the previous subsection help us in understanding the behavior of crowdsourcing systems. We can define an ordering principle for quality of crowds in terms of the quality of their distributed inference performance. This is a valuable concept since it provides us a tool to evaluate a given crowd. Such a valuation could be used by the task manager to pick the appropriate crowd for the task based on the performance requirements. For example, if the task manager is interested in constraining the misclassification probability of his/her task to $\epsilon$ while simultaneously minimizing the required crowd size, the above expressions can be used to choose the appropriate crowd.

\begin{theorem}[Ordering of Crowds]
\label{theorem_ordering}
Consider crowdsourcing systems involving crowd $\mathcal{C}(\mu)$ of workers with i.i.d.\ reliabilities with mean $\mu$. 
Crowd $\mathcal{C}(\mu)$ performs better than crowd $\mathcal{C}(\mu')$ for classification if and only if $\mu > \mu'$.
\end{theorem}
\begin{IEEEproof}
Follows since average misclassification probabilities depend only on the mean of the reliabilities of the crowd. 
\end{IEEEproof}
Since the performance criterion is average misclassification probability, this can be regarded as a weak criterion of crowd-ordering in the mean sense. Thus, with this crowd-ordering, better crowds yield better performance in terms of average misclassification probability.  Indeed, misclassification probability decreases with better quality crowds. In this paper, the term reliability has been used to describe the individual worker's reliability while the term quality is a description of the total reliability of a given crowd (a function of mean $\mu$ of worker reliabilities). For example, for the spammer-hammer model, quality of the crowd is a function of the number of hammers in the crowd, while the individual crowd workers have different reliabilities depending on whether the worker is a spammer or a hammer.

\begin{proposition}
\label{lemma_quality}
Average misclassification probability reduces with increasing quality of the crowd.
\end{proposition}
\begin{IEEEproof}
Follows from Props.~\ref{prop1} and \ref{prop:maj_er_ind} for coding- and majority-based approaches, respectively.
\end{IEEEproof}

To get more insight, we simulate a crowdsourcing system with coding as follows: $N=10$ workers take part in a classification 
task with $M=4$ equiprobable classes. A good code matrix $\mathbf{A}$ is found by simulated annealing \cite{WangHVC2005}:

\begin{equation}
\mathbf{A}=[5,12,3,10,12,9,9,10,9,12] \mbox{.}
\label{eq:first_code}
\end{equation}
Here and in the sequel, we represent code matrices as a vector of $M$ bit integers. Each integer $r_j$ represents a column 
of the code matrix $\mathbf{A}$ and can be expressed as $r_j=\sum_{l=0}^{M-1}a_{lj}\times 2^l$. For example, the 
integer $5$ in column $1$ of $\mathbf{A}$ represents $a_{01}=1$, $a_{11}=0$, $a_{21}=1$ and $a_{31}=0$.

\begin{figure}
  \centering
  \includegraphics[width = 3.5in, height=!]{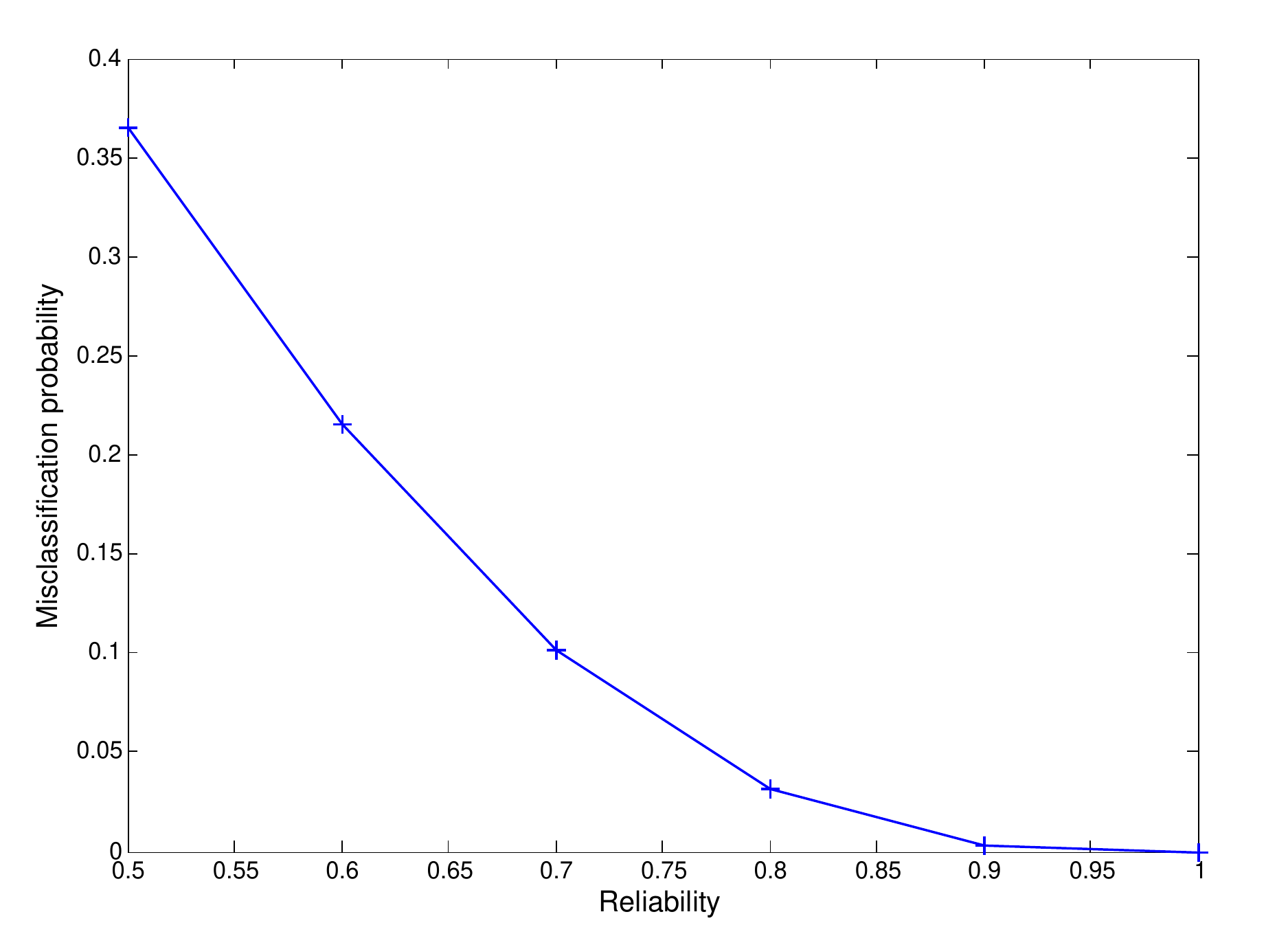}
  \caption{Coding-based crowdsourcing system misclassification probability as a function of worker reliability.}
  \label{err_rel}
\end{figure}

Let us look at the setting where all the workers have the same reliability $p_j=p$. Fig.~\ref{err_rel} shows the probability 
of misclassification as a function of $p$.  As is apparent, the probability of misclassification reduces with reliability 
and approaches 0 as $p \to 1$, as expected.

\begin{figure}
\centering
\includegraphics[width = 3.5in, height=!]{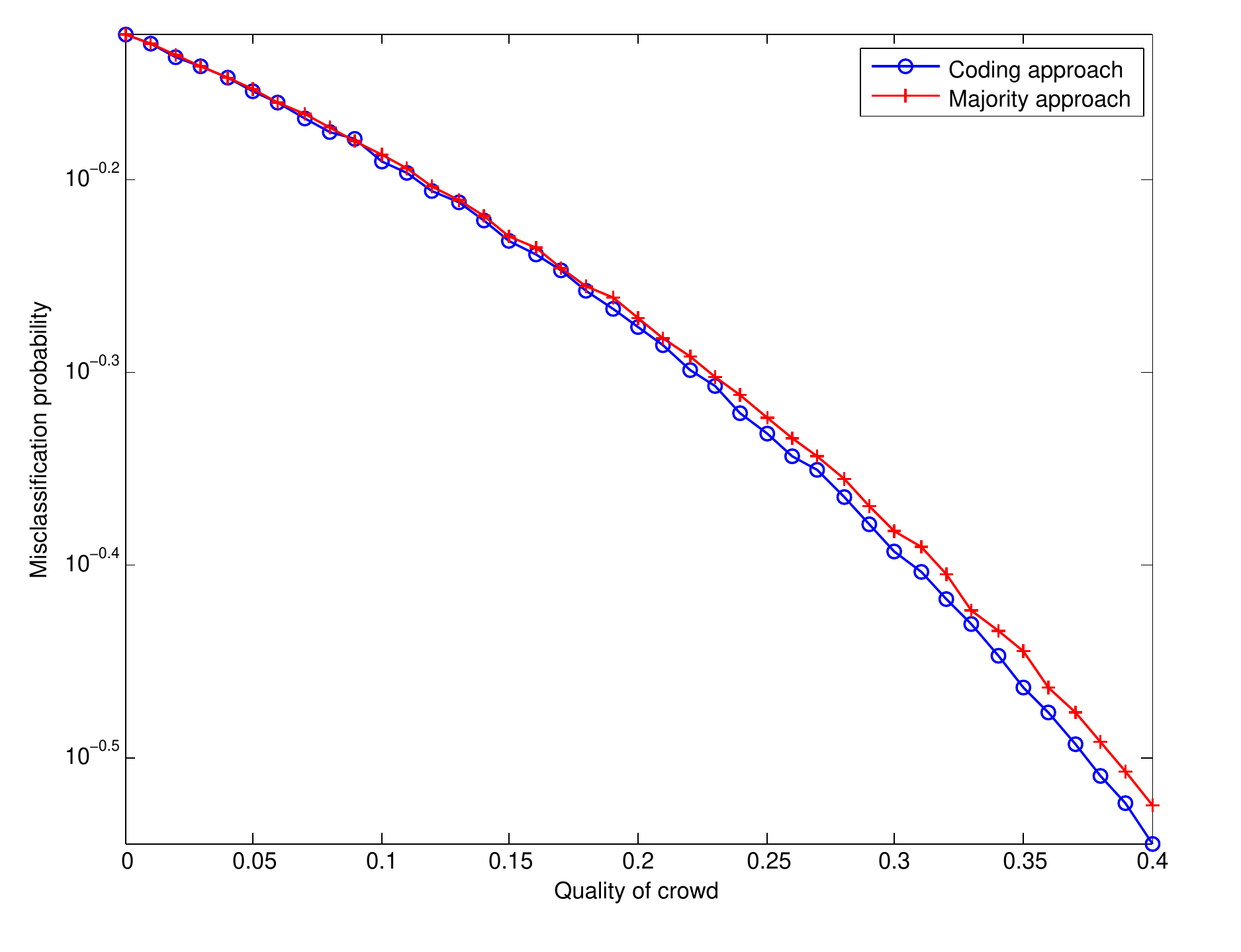}
\caption{Misclassification probability as a function of crowd quality using coding- and majority-based approaches with the spammer-hammer model, $(M=4, N=10)$.}
\label{err_quality_majority}
\end{figure}

Now we compare the performance of the coding-based approach to the majority-based approach. Fig.~\ref{err_quality_majority} shows 
misclassification probability as a function of crowd quality for $N=10$ workers taking part in an $(M=4)$-ary classification task. 
The spammer-hammer model, where spammers have reliability $p=1/M$ and hammers have reliability $p=1$, is used. The figure shows a slight improvement in performance over majority vote when code matrix \eqref{eq:first_code} is used.

\begin{figure}
\centering
\includegraphics[width = 3.5in, height=!]{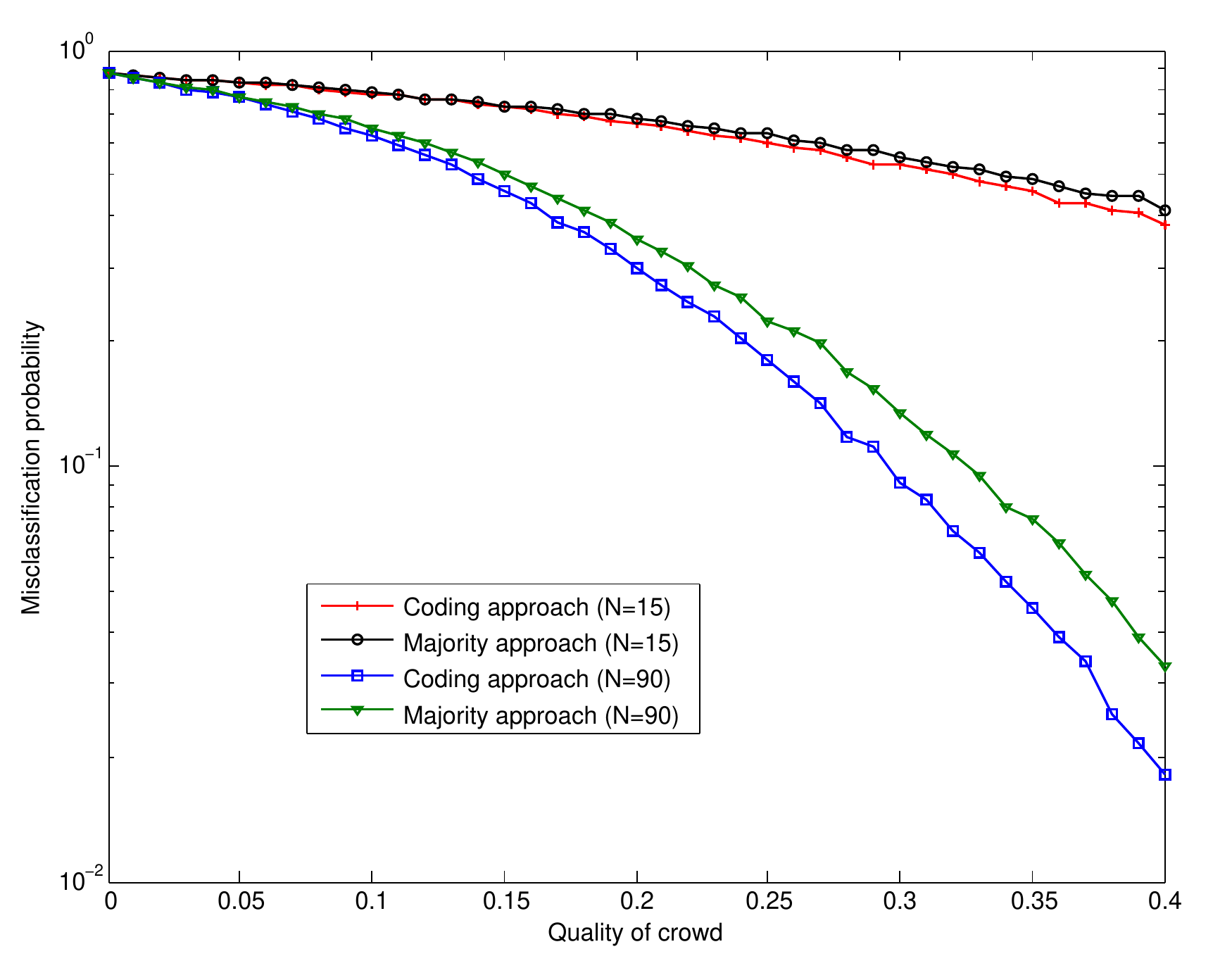}
\caption{Misclassification probability as a function of crowd quality using coding- and majority-based approaches with the spammer-hammer model, $(M=8)$.}
\label{err_quality_majority_inc}
\end{figure}

\begin{figure}
\centering
\includegraphics[width = 3.5in,height=!]{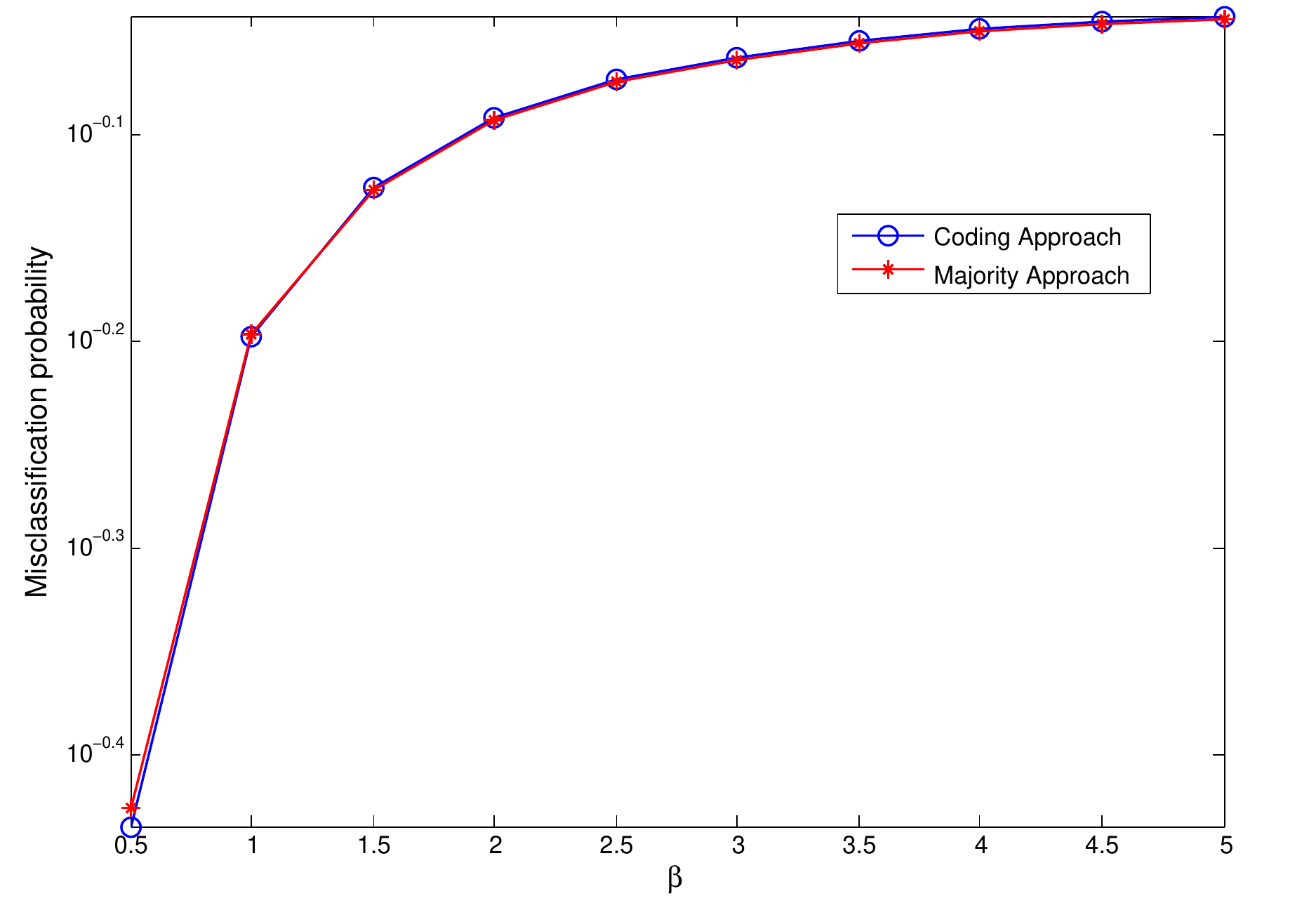}
\caption{Misclassification probability as a function of $\beta$ using coding- and majority-based approaches with the Beta($\alpha=0.5$, $\beta$) model, $(M=4, N=10)$.}
\label{err_quality_majority_beta}
\end{figure}

\begin{figure}
\centering
\includegraphics[width = 3.5in, height=!]{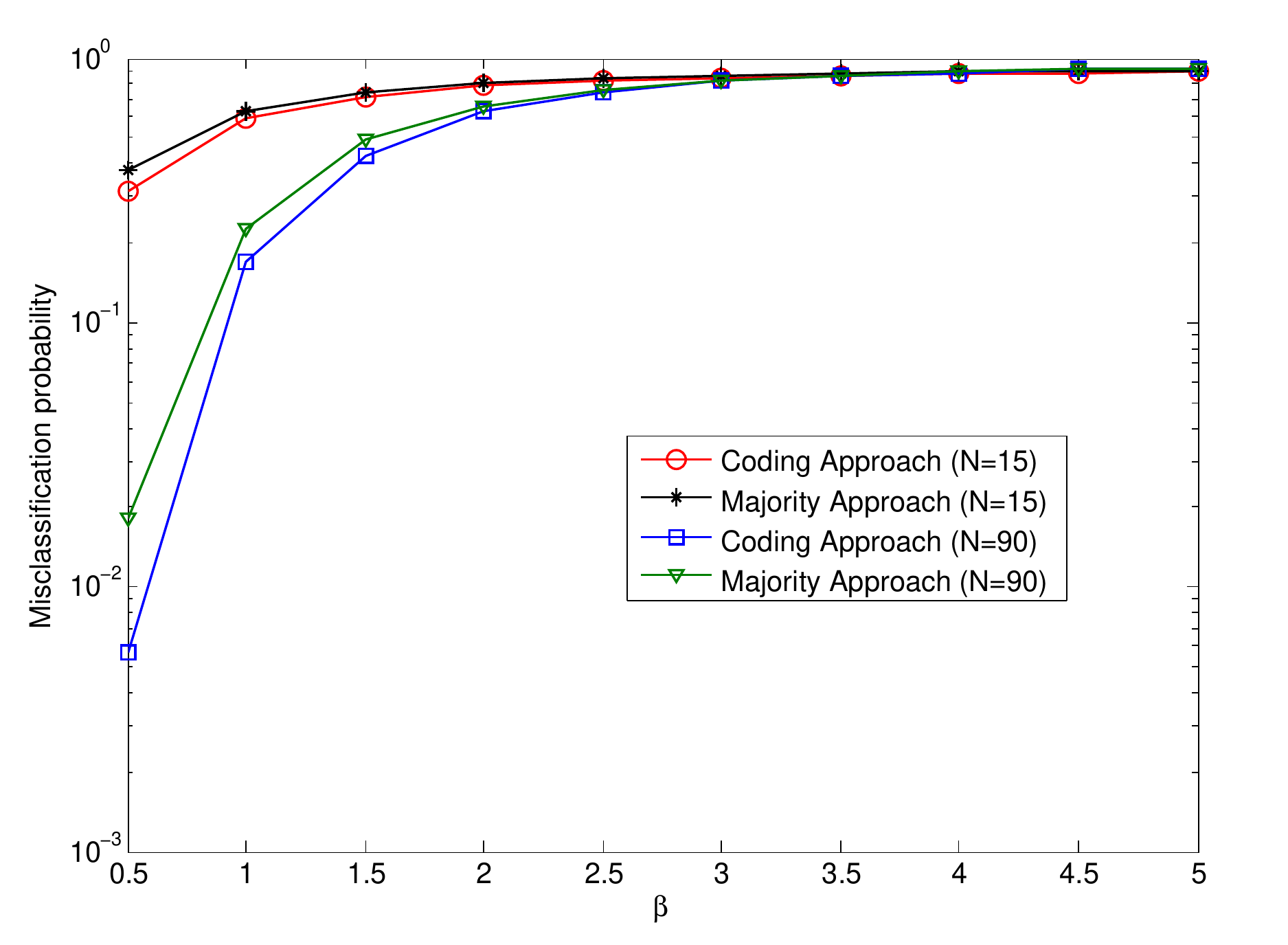}
\caption{Misclassification probability as a function of $\beta$ using coding- and majority-based approaches with the Beta($\alpha=0.5$, $\beta$) model, $(M=8)$.}
\label{err_quality_majority_beta_inc}
\end{figure}

\begin{figure}
\centering
\includegraphics[width = 3.5in, height=!]{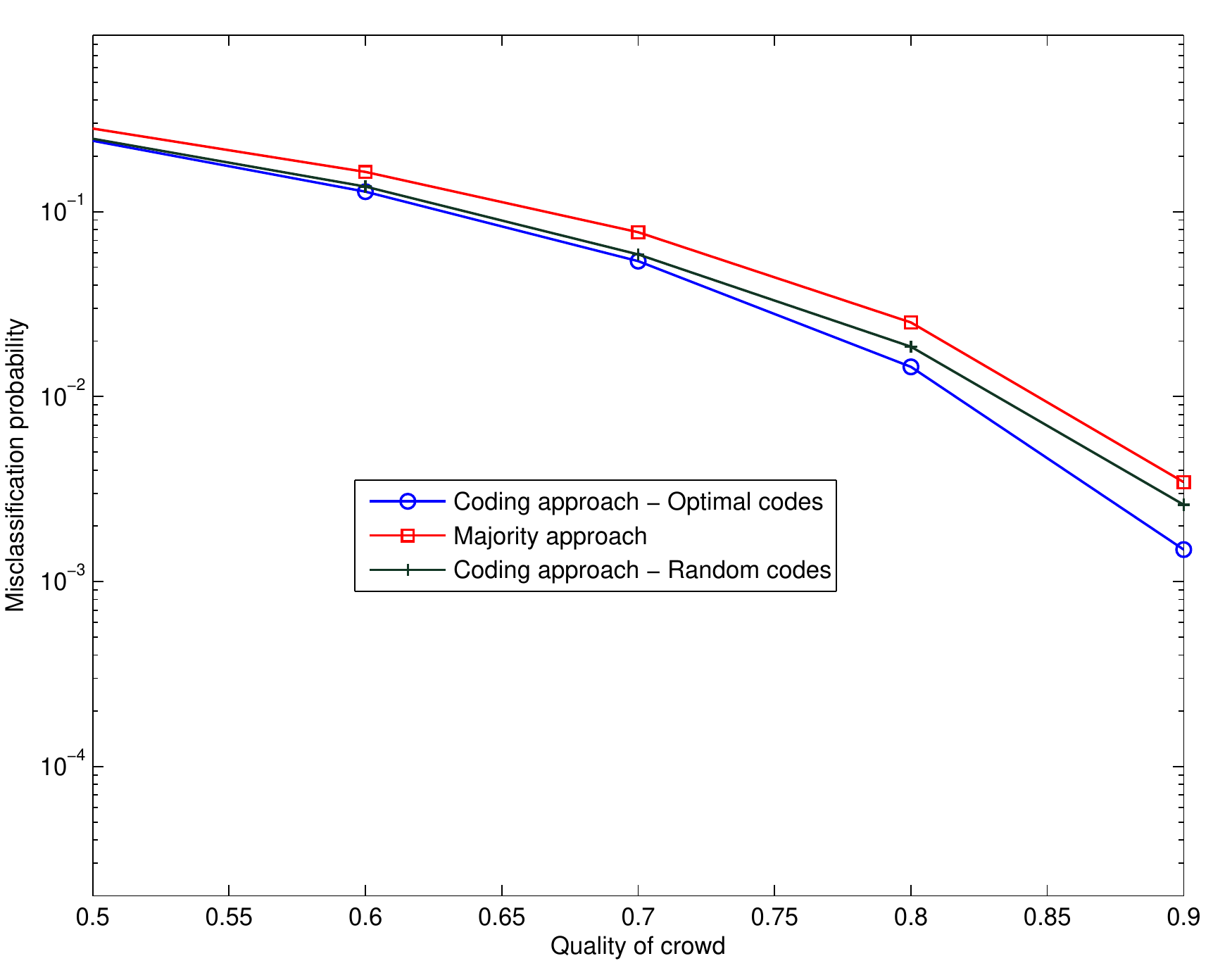}
\caption{Misclassification probability as a function of crowd quality using optimal code matrix, random code matrix for coding-based approach and majority approach with the spammer-hammer model, $(M=8, N=15)$.}
\label{fig:error_prob_8_random}
\end{figure}

We consider a larger system with increased $M$ and $N$. A good code matrix $\mathbf{A}$ for $N=15$ and $M=8$ is found by cyclic column replacement:
\begin{equation}
\mathbf{A} = [150,150,90,240,240,153,102,204,204,204,170,170,170,170,170] \mbox{.}
\label{eq:codematrix_8}
\end{equation}
The code matrix for the system with $N=90$ and $M=8$ is formed sub-optimally by concatenating the columns of \eqref{eq:codematrix_8} six times. Due to the large system size, it is computationally very expensive to optimize for the code matrix using either the simulated annealing or cyclic column replacement methods. Therefore, we concatenate the columns of \eqref{eq:codematrix_8}. This can be interpreted as a crowdsourcing system of 90 crowd workers consisting of 6 sub-systems with 15 workers each which are given the same task and their data is fused together. In the extreme case, if each of these sub-systems was of size one, it would correspond to a majority vote where all the workers are posed the same question. Fig.~\ref{err_quality_majority_inc} shows the performance when $M=8$ and $N$ takes the two values: $N=15$ and $N=90$.  These figures suggest that the gap in performance generally increases for larger system size.
Similar observations hold for the beta model of crowds, see Figs.~\ref{err_quality_majority_beta} and \ref{err_quality_majority_beta_inc}.
Good codes perform better than majority vote as they diversify the binary questions which are asked to the workers. From extensive simulation results, we have found that the coding-based approach is not very sensitive to the choice of code matrix $\mathbf{A}$ as long as we have approximately equal number of ones and zeroes in every column. However, if we use any code randomly, performance may degrade substantially, especially when the quality of crowd is high. For example, consider a system consisting of $N=15$ workers performing a $(M=8)$-ary classification task. Their reliabilities are drawn from a spammer-hammer model and Fig.~\ref{fig:error_prob_8_random} shows the performance comparison between coding-based approach using the optimal code matrix, majority-based approach and coding-based approach using a random code matrix with equal number of ones and zeroes in every column. We can observe that the performance of the coding-based approach with random code matrix deteriorates for higher quality crowds.

\subsubsection{Experimental Results}
\label{sec:real}
In this section, we test the proposed coding- based approach on six publicly available Amazon Mechanical Turk data sets---quantized 
versions of the data sets in \cite{SnowOJN2008}: the anger, disgust, fear, joy, sadness and surprise datasets of the affective text task. 
Each of the data sets consist of $100$ tasks with $N=10$ workers taking part in each. Each worker reports 
a value between $0$ and $100$, and there is a gold-standard value for each task. For our analysis, we quantize values by dividing 
the range into $M=8$ equal intervals.  We compare the majority -based approach with our proposed coding-based  approach. 
A good optimal code matrix for $N=10$ and $M=8$ is designed by simulated annealing \cite{WangHVC2005}:
\begin{equation}
\mathbf{A}=[113,139,226,77,172,74,216,30,122] \mbox{.}
\end{equation}

Table~\ref{my_table_iid} compares the performance of the coding- and majority-based approaches. 
The values in Table~\ref{my_table_iid} are the fraction of wrong decisions made, as compared with the gold-standard value. As indicated, 
the coding-based approach performs at least as well as the majority-based approach in 4 of 6 cases considered.  We expect the gap in performance to increase as 
problem size $M$ and crowd size $N$ increase. Also, while it is true that the coding-based approach is only slightly better than the majority approach in the cases considered in Table~\ref{my_table_iid}, this comparison only shows the benefit of the proposed coding-based approach in terms of the fusion scheme. The datasets contain data of tasks where the workers have reported continuous values and, therefore, it does not capture the benefit of asking binary questions. This aspect is a major benefit of the proposed coding-based approach whose empirical testing we defer to future work. 

\begin{table}
\begin{center}
\caption{Fraction of errors using coding- and majority-based approaches}
\begin{tabular}{|l|l|l|}
\hline
	Dataset & Coding-based approach  & Majority-based approach\\
   \hline 
   \hline
   Anger & 0.31 & 0.31\\ 
   \hline
   Disgust & 0.26 & 0.20\\ 
   \hline
   Fear & 0.32 & 0.30\\ 
   \hline
   Joy & 0.45 & 0.47\\ 
   \hline
   Sadness & 0.37 & 0.39\\ 
   \hline
   Surprise & 0.59 & 0.63\\ 
   \hline
\end{tabular}
\label{my_table_iid}
\end{center}
\end{table}

\section{Crowdsourcing System with Peer-dependent Reward Scheme}
\label{sec:peer}
In this section, we consider the crowdsourcing system wherein the crowd workers are paired into groups of two and their reward value is based on the comparative performance among the paired workers \cite{HuangF2013}.  This has been proposed as a method for increasing worker motivation.

\subsection{Model}
We consider two kinds of worker pairings: competitive and teamwork. Both kinds can be captured by considering crowd reliability models where worker reliabilities are no longer independent as in Sec.~\ref{sec:iid}. For simplicity, let us assume that $N$ is even so that each worker $j$ has a corresponding partner ${j_p}$ and there are a total of $N/2$ pairs. The reliabilities of these paired workers are correlated with covariance $\rho$, which is assumed to be the same for all pairs due to the identical nature of workers. Also, workers within a pair are independent of the workers outside their pair. Hence:
\begin{equation}
\label{eq:cov}
cov(p_i,p_j)=
\begin{cases}
0,	& \text{if $i\neq j_p$}\\
\rho, &\text{if $i = j_p$} \mbox{.}
\end{cases}
\end{equation}

The value of $\rho$ depends on whether workers are paired for teamwork or for competition \cite{HuangF2013}. An $(N,M,\mu,\rho)$ crowdsourcing system has $N$ workers performing an $M$-ary classification task and having reliabilities $p_j$ which are identical random variables with mean $\mu$ and covariance structure defined by \eqref{eq:cov}. 

\subsection{Performance Characterization}
In this section, we derive the misclassification probability when the crowd workers are correlated. As described above, this scenario takes place when the workers are paired with each other. 

\begin{proposition}
\label{prop:coding_pdr}
Consider an $(N,M,\mu,\rho)$ crowdsourcing system. The expected misclassification probability using code matrix $\mathbf{A}$ is:
\begin{multline}
\label{eq:err_coding_pdr}
P_e(\mu,\rho)=\frac{1}{M}\sum_{\underline{i},l}C_{\underline{i}}^l\prod_{j=1}^{\frac{N}{2}}
\bigg[(1-i_j)(1-i_{j_p})+(1-i_j)(2i_{j_p}-1)\left(\mu a_{lj_p}+\frac{(1-\mu)}{(M-1)}\sum_{k\neq l}a_{kj_p}\right)+(1-i_{j_p})(2i_{j}-1)\\
\left(\mu a_{lj}+\frac{(1-\mu)}{(M-1)}\sum_{k\neq l}a_{kj}\right)+(2i_j-1)(2i_{j_p}-1)\bigg((\rho +\mu^2)a_{lj}a_{lj_{p}}+\frac{\mu-(\rho+\mu^2)}{(M-1)}\left(a_{lj}\sum_{k\neq l}a_{kj_p}+a_{lj_p}\sum_{k\neq l}a_{kj}\right)\\
+\frac{4r}{M^2}\sum_{k\neq l}a_{kj_p}\sum_{k\neq l}a_{kj}\bigg)\bigg]
\end{multline}
where $r=\left(\frac{M}{2(M-1)}\right)^2[(1-\mu)^2+\rho]$, $\underline{i}=[i_1, \cdots, i_N] \in \{0, 1\}^N$ is the received codeword and $C^l_{\underline{i}}$ is the cost associated with a global decision $H_l$ when the received vector is $\underline{i}$. This cost is given in \eqref{eq:cost}.
\end{proposition}

\begin{IEEEproof}
Let $P_{e,\underline{p}}$ denote the misclassification probability given the reliabilities of the $N$ workers. Then, if $u_j$ denotes the 
bit sent by the worker $j$ and the global decision is made using the Hamming distance criterion,
\begin{equation}
P_{e,\underline{p}}=\frac{1}{M}\sum_{\underline{i},l}P(\underline{u}=\underline{i}|H_l)C_{\underline{i}}^l\mbox{.}
\end{equation}
Since local decisions are conditionally independent, $P(\underline{u}=\underline{i}|H_l)=\prod_{j=1}^N P(u_j=i_j|H_l)$.
Further, 
\begin{align*}
P(u_j=i_j|H_l) &= i_jP(u_j=1|H_l)+(1-i_j)P(u_j=0|H_l)\\
&=(1-i_j)+(2i_j-1)P(u_j=1|H_l)\\
&=(1-i_j)+(2i_j-1)\sum_{k=1}^Ma_{kj}P(y_j=k|H_l)\\
&=(1-i_j)+(2i_j-1)\Bigg(p_j a_{lj}+\frac{(1-p_j)}{(M-1)}\sum_{k\neq l}a_{kj}\Bigg)
\end{align*}
where $y_j$ is the local decision made by worker $j$. Note that the reliabilities $p_j$ of workers across pairs are independent while the workers within the pair are correlated according to \eqref{eq:cov}. Therefore, we get:
\begin{align}
\label{eq:correlated_probs}
&E\left[\prod_{j=1}^N P(u_j=i_j|H_l)\right] \notag\\
&= \prod_{j=1}^{N/2}E\left[P(u_j=i_j|H_l)P(u_{j_p}=i_{j_p}|H_l)\right]\notag \\
&= \prod_{j=1}^{N/2}E\Bigg[\left((1-i_j)+\left(p_j a_{lj} + \tfrac{(1-p_j)}{(M-1)}\sum_{k\neq l}a_{kj}\right)(2i_j-1)\right)\left((1-i_{j_p})+\left(p_{j_p} a_{lj_p} + \tfrac{(1-p_{j_p})}{(M-1)}\sum_{k\neq l}a_{kj_p}\right)(2i_{j_p}-1)\right)\Bigg]
\end{align}
The above equation, correlation structure \eqref{eq:cov} and definition $r=\left(\frac{M}{2(M-1)}\right)^2[(1 - \mu)^2 + \rho]$ yield the desired result.
\end{IEEEproof}

\subsection{Majority Voting}
As mentioned before, for the sake of comparison, we derive error performance expressions for the majority-based approach too. Consider majority vote, with $N$ divisible by $2\log_2M$.

\begin{proposition}
\label{prop:maj_pdr}
Consider an $(N,M,\mu,\rho)$ crowdsourcing system. The expected misclassification probability using majority rule is: 
\begin{equation}
\label{eq:err_majority_pdr}
P_e(\mu,\rho)=1-\frac{1}{M}\Bigg[1+\sum_{j=\lfloor\frac{\tilde{N}}{2}+1\rfloor}^{\tilde{N}}b_j(\tilde{N},q,r)\left[(1-2q+r)^{(j-\frac{\tilde{N}}{2})}-r^{(j-\frac{\tilde{N}}{2})}\right] \Bigg]^{\log_2M},
\end{equation}
where 
\begin{equation}
b_j(\tilde{N},q,r)=\sum_{g=0}^{\lfloor\frac{\tilde{N}-j}{2}\rfloor}\binom{\frac{\tilde{N}}{2}}{g} \binom{\frac{\tilde{N}}{2}-g}{j+g-\frac{\tilde{N}}{2}}\left[2\left(q-r\right)\right]^{(\tilde{N}-j-2g)}(r-2qr+r^2)^g,
\end{equation}
$\tilde{N}=\frac{N}{\log_2M}$, $q=\frac{M(1-\mu)}{2(M-1)}$, and $r=\left(\frac{M}{2(M-1)}\right)^2[(1-\mu)^2+\rho]$.
\end{proposition}

\begin{IEEEproof}
In a majority-based approach, $\tilde{N}=\tfrac{N}{\log_2M}$ workers send information regarding the $i$th bit of their local decision, $i=1,\ldots,\log_2 M$. For a correct global decision, all bits have to be correct. Consider the $i$th bit and let $P^i_{c,\underline{p}}$ be the probability of the $i$th bit being correct given the reliabilities of the $\tilde{N}$ workers sending this bit. Also, assume that the paired workers send the same bit information. Then,
\begin{equation}
\label{eq:Pc_pdr}
P^i_{c,\underline{p}}=\frac{P_d +1 - P_f}{2} \mbox{,}
\end{equation}
where $P_d$ is the probability of detecting the $i$th bit as `1' when the true bit is `1' and $P_f$ is the probability of detecting the $i$th bit as `1' when the true bit is `0'. Note that `0' and `1' are equiprobable since all the hypotheses are equiprobable. Under majority rule for this $i$th bit,
\begin{equation}
\label{eq:pd_pdr}
P_d=\sum_{j=\lfloor\frac{\tilde{N}}{2}+1\rfloor}^{\tilde{N}}\sum_{\forall G_j}\prod_{k\in G_j}\left(1-\frac{M(1-p_k)}{2(M-1)}\right)\prod_{k \notin G_j} \frac{M(1-p_k)}{2(M-1)},
\end{equation}
where $G_j$ is a set of $j$ out of $\tilde{N}$ workers who send bit value `1' and $\frac{M(1-p_k)}{2(M-1)}$ is the probability of the $k$th worker making a wrong decision for the $i$th bit. Similarly,
\begin{equation}
\label{eq:pf_pdr}
P_f=\sum_{j=\lfloor\frac{\tilde{N}}{2}+1\rfloor}^{\tilde{N}}\sum_{\forall G_j}\prod_{k\in G_j}\frac{M(1-p_k)}{2(M-1)}\prod_{k \notin G_j}\left(1-\frac{M(1-p_k)}{2(M-1)}\right),
\end{equation}

Now the overall probability of correct decision is given by 
\begin{equation}
P_{c,\underline{p}}=\prod_{i=1}^{\log_2M}P_{c,\underline{p}}^i.
\end{equation}

Since the workers across the groups are independent, the expected probability of correct decision $P_c$ is:
\begin{equation}
\label{eq:prod_pdr}
P_c = \prod_{i=1}^{\log_2M}E[P^i_{c,\underline{p}}],
\end{equation} 
where expectation is with respect to $\underline{p}$. Within the same group, paired workers exist who are correlated and, therefore, the reliabilities are not independent. Let $g$ pairs of workers be present in the set $G_j^c$, where $0\leq g\leq\frac{\tilde{N}-j}{2}$. This implies that $2g$ workers in $G_j^c$ are correlated with their partners in the same group and the remaining $(\tilde{N}-j-2g)$ are correlated with their paired workers in $G_j$. Therefore, there are $(j+g-\frac{\tilde{N}}{2})$ pairs of correlated workers in $G_j$. The number of such divisions of workers into $G_j$ and $G_j^c$ such that there are exactly $g$ pairs of correlated workers in $G_j^c$ is determined as the number of ways of choosing $g$ possible pairs from a total of $\frac{\tilde{N}}{2}$ pairs for $G_j^c$ and $(j+g-\frac{\tilde{N}}{2})$ pairs of correlated workers from the remaining $(\frac{\tilde{N}}{2}-g)$ pairs for $G_j$. The remaining $(\tilde{N}-j-2g)$ workers of $G_j^c (G_j)$ have their paired workers in $G_j (G_j^c)$ and this gives the following number of ways of such divisions: 
\begin{equation}
\label{eq:number}
N_g=\binom{\frac{\tilde{N}}{2}}{g} \binom{\frac{\tilde{N}}{2}-g}{j+g-\frac{\tilde{N}}{2}}2^{(\tilde{N}-j-2g)}.
\end{equation}

Define $q=\frac{M(1-\mu)}{2(M-1)}$ as the expected probability of a worker deciding a wrong bit and 
\begin{eqnarray}
r=E\left[\frac{M(1-p_k)}{2(M-1)}\frac{M(1-p_{k_p})}{2(M-1)}\right]\\
=\left(\frac{M}{2(M-1)}\right)^2[(1-\mu)^2+\rho]
\end{eqnarray}
as the expected probability that both the workers in a correlated worker pair send wrong bit information. Taking the expectation of $P_d$ and $P_f$ (cf.~\eqref{eq:pd_pdr}, \eqref{eq:pf_pdr}) and using \eqref{eq:number}, we get:
\begin{eqnarray}
E[P_d]&=&\sum_{j=\lfloor\frac{\tilde{N}}{2}+1\rfloor}^{\tilde{N}}\sum_{g=0}^{\lfloor\frac{\tilde{N}-j}{2}\rfloor}\binom{\frac{\tilde{N}}{2}}{g} \binom{\frac{\tilde{N}}{2}-g}{j+g-\frac{\tilde{N}}{2}}\nonumber\\
&&\left[2\left(q-r\right)\right]^{(\tilde{N}-j-2g)}r^g(1-2q+r)^{(j+g-\frac{\tilde{N}}{2})}\nonumber\\
&=&\sum_{j=\lfloor\frac{\tilde{N}}{2}+1\rfloor}^{\tilde{N}}b_j(\tilde{N},q,r)(1-2q+r)^{(j-\frac{\tilde{N}}{2})}
\label{eq:Pd_expected_pdr}
\end{eqnarray}
and
\begin{eqnarray}
E[P_f]&=&\sum_{j=\lfloor\frac{\tilde{N}}{2}+1\rfloor}^{\tilde{N}}\sum_{g=0}^{\lfloor\frac{\tilde{N}-j}{2}\rfloor}\binom{\frac{\tilde{N}}{2}}{g} \binom{\frac{\tilde{N}}{2}-g}{j+g-\frac{\tilde{N}}{2}}\nonumber\\
&&\left[2\left(q-r\right)\right]^{(\tilde{N}-j-2g)}r^{(j+g-\frac{\tilde{N}}{2})}(1-2q+r)^g\nonumber\\
&=&\sum_{j=\lfloor\frac{\tilde{N}}{2}+1\rfloor}^{\tilde{N}}b_j(\tilde{N},q,r)r^{(j-\frac{\tilde{N}}{2})}\mbox{.}
\label{eq:Pf_expected_pdr}
\end{eqnarray}
Using \eqref{eq:Pc_pdr}, \eqref{eq:prod_pdr}, \eqref{eq:Pd_expected_pdr}, and \eqref{eq:Pf_expected_pdr}, we get the desired result.
\end{IEEEproof}

We can make some observations.  First, when workers are not paired and instead perform individually, they are independent and $\rho=0$. Therefore, $r=q^2$ and $P_e(\mu,0)$ for majority- and coding-based approaches from Props.~\ref{prop:coding_pdr} and \ref{prop:maj_pdr} reduce to the results from Sec.~\ref{sec:iid}.  Second, a crowd with mean $\mu=1/M$, i.e., a crowd with workers of poor reliabilities on an average, has $q=1/2$ and performs no better than random classification which gives $P_c(1/M,\rho)=1/M$ for the majority case or $P_e(1/M,\rho)=\frac{(M-1)}{M}$. Similar observations can be made for the coding-based approach.

\subsection{Performance Evaluation}
\begin{figure}
\centering
\includegraphics[width = 3.5in,height=!]{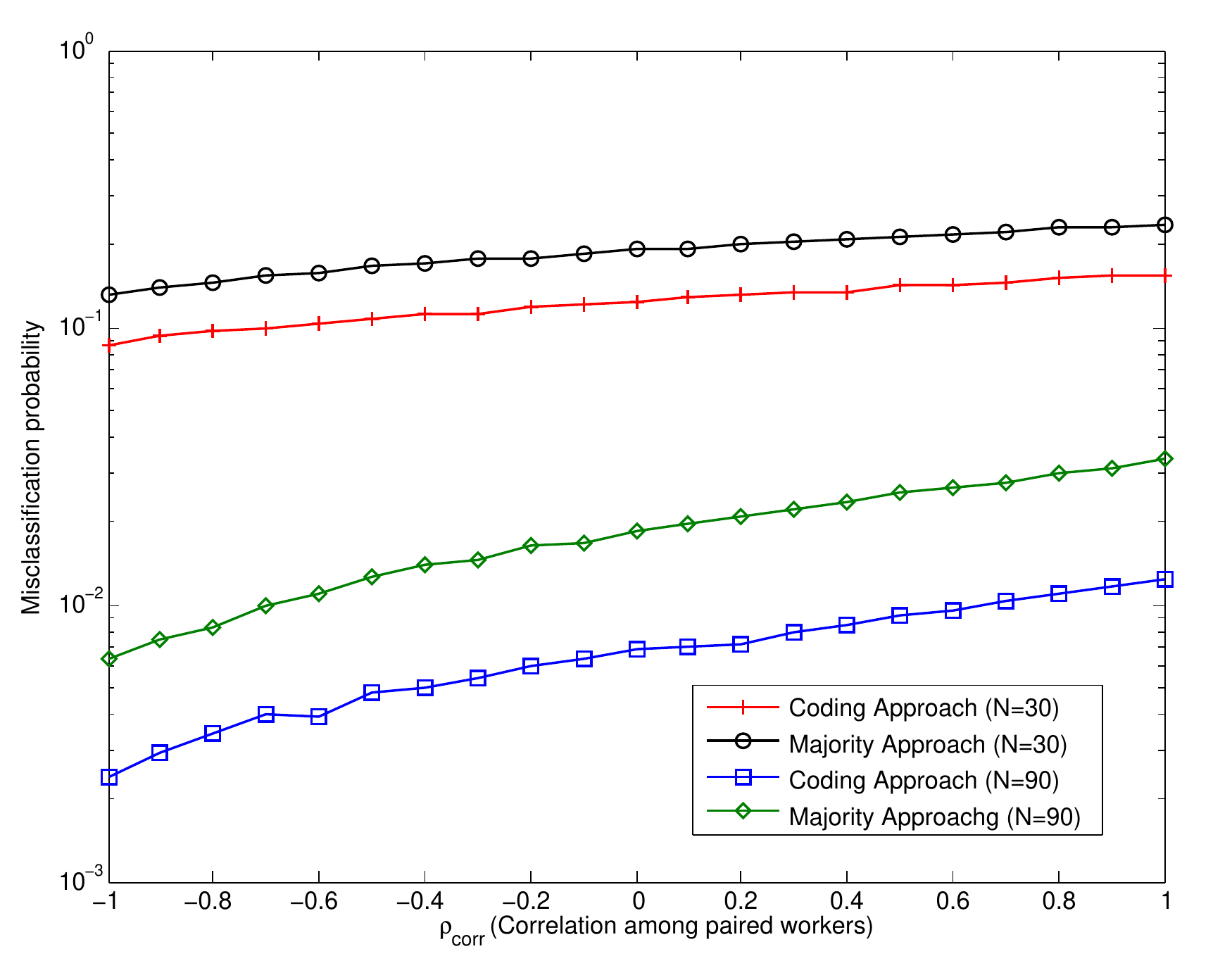}
\caption{Misclassification probability as a function of $\rho_{corr}$ using coding- and majority-based approaches with the Beta($\alpha=0.5$, $\beta=0.5$) model, $(M=8)$.}
\label{fig:peer_8}
\end{figure}

Next, we evaluate the performance of this system with peer-dependent reward scheme. As mentioned before, such a scheme results in correlation among the crowd workers. Fig.~\ref{fig:peer_8} shows the performance of a system with peer-dependent reward scheme with varying correlation parameter ($\rho_{corr}$). Note that the plots are with respect to the correlation coefficient ($\rho_{corr}$) and the equations are with respect to the covariance parameter ($\rho$). The plots are for a system with $N$ crowd workers and $M=8$, i.e., performing $8$-ary classification. As the figure suggests, the performance of the system improves when the workers are paired in the right manner. Note that when the crowdsourcing system with peer-dependent reward scheme has $\rho_{corr} =0$, it reduces to the system with individual crowd workers considered in Sec. \ref{sec:iid}. Fig. \ref{fig:peer_8} includes the system with independent crowd workers as a special case when $\rho_{corr} =0$. Also, these results suggest that having a negative correlation among workers results in better performance while a positive correlation deteriorates system performance. This observation can be attributed to the amount of diversity among the crowd workers. When $\rho_{corr}=1$, the workers are correlated with each other while a correlation value of $\rho_{corr}=-1$ corresponds to a diverse set of workers which results in improved performance.

\section{Crowdsourcing System with Common Sources of Information}
\label{sec:dep}
In this section, we generalize our model further by considering dependence among the observations of the crowd workers; previous sections had considered independent local observations by crowd workers. Although Sec.~\ref{sec:peer} considered paired workers with correlated reliabilities, here observations by crowd workers are dependent too.  Crowd workers may have dependent observations when they share a common information source \cite{QiAHH2013}.

\subsection{Model}
One way to capture dependence among workers' observations is to assume that there is a set of latent groups $\{S_1, S_2,\cdots\}$, and each crowd worker $j$ is assigned to one group $S_{s_{j}}$ where $s_j \in \{1,2,\cdots\}$ is a random variable indicating membership. Each group $S_l$ is modeled to have an associated group reliability $r_l \in [0,1]$ which determines the general reliability of the group. When a group represents the set of crowd workers who share a common source of information, the reliability $r_l$ of the group $S_l$ represents the reliability of the information source. Associated with every crowd worker $j$, then, is the worker's reliability $p_j \in [0,1]$ which determines the worker's reliable use of the information.

We now describe a generative process for this model, the Multi-Source Sensing Model described in \cite{QiAHH2013}. 
\begin{enumerate}
\item Draw $\underline{\lambda} \sim \text{GEM}(\kappa)$, i.e., stick breaking construction with concentration $\kappa$.
For this, the stick-breaking construction GEM($\kappa$) (named after Griffiths, Engen and McCloskey) discussed in \cite{QiAHH2013} is used. Specifically, in GEM($\kappa$), a set of random variables $\underline{\gamma}=\{\gamma_1,\gamma_2, \cdots\}$ are independently drawn from the beta distribution $\gamma_i~\sim~\beta(1,\kappa)$. They define the mixing weights $\underline{\lambda}$ of the group membership component such that $P(s_j=l|\underline{\gamma})=\lambda_l=\gamma_l\prod_{g=0}^{l-1}(1-\gamma_g)$. 

\item For each worker $j$, draw its group assignment $s_j|\underline{\lambda} \sim \text{Discrete}(\underline{\lambda})$, 
where $s_j|\underline{\lambda} \sim \text{Discrete}(\underline{\lambda})$ denotes a discrete distribution, which generates the value $s_j = l$ with probability $\lambda_l$.

\item For each group $S_l$, draw its group reliability $r_l \sim F(\mu)$, 
where $F(\mu)$ is the group reliability distribution with mean $\mu$. Some possible examples are the beta model where reliabilities are beta distributed or the spammer-hammer model.
\item For each crowd worker $j$, draw their reliability $p_j \sim \beta\left(\frac{r_{s_j}}{1-r_{s_j}},1\right)$ such that $E[p_j]=r_{s_j}$
\item For each crowd worker $j$, draw his observation $y_j$ based on reliability $p_j$ and true hypothesis $H_l$ (cf. \eqref{eq:obs_model})
\end{enumerate}

Learning does not require prior knowledge on the number of groups, due to the stick-breaking process, but note the dependence among observations depends on the number of latent groups which itself is a function of the concentration parameter $\kappa$. Due to grouping of workers, reliabilities may not be independent random variables as in Sec.~\ref{sec:peer}. An $(N, M, \mu, \kappa)$ crowdsourcing system consists of $N$ grouped, but unpaired crowd workers performing $M$-ary classification, whereas an $(N, M, \mu, \rho, \kappa)$ crowdsourcing system has $N$ grouped and paired crowd workers with pairing covariance $\rho$ performing $M$-ary classification.

\subsection{Performance Characterization}

We first consider crowd workers grouped by latent variables, but without peer-dependent rewards.

\begin{proposition}
\label{prop:error_prob_dep_uncorr}
Consider an $(N, M, \mu, \kappa)$ crowdsourcing system. The expected misclassification probability using code matrix $\textbf{A}$ is:
\begin{eqnarray}
&&P_e(\mu, \kappa)=\frac{1}{M}\sum_{\underline{i}, l, \underline{s}}C_{\underline{i}}^l P(\underline{S}=\underline{s}|H_l)\prod_{j=1}^N \left[\left(\mu a_{lj} + \frac{(1-\mu)}{(M-1)}\sum_{k\neq l}a_{kj}\right)(2i_j-1)+(1-i_j)\right],\nonumber
\end{eqnarray}
where $\underline{i}=[i_1,\cdots,i_N] \in \{0,1\}^N$ is the received codeword, $\underline{s}=[s_1,\cdots,s_N] \in \{0,\cdots, L-1\}^N$ is the group assignment, $L$ is the number of groups and $C_{\underline{i}}^l$ is the cost associated with a global decision $H_l$ when the received vector is $\underline{i}$. This cost is given in \eqref{eq:cost}. 
The probability $P(\underline{S}=\underline{s}|H_l)$ is a function of the concentration parameter $\kappa$ and is:
\begin{equation}
P(\underline{S}=\underline{s}|H_l)=\frac{1}{[\beta(1,\kappa)]^L}\prod_{l=0}^{L-1}\beta\left(n_l+1, N+\kappa-\sum_{g=0}^l n_g\right)
\end{equation}
where $\beta(\cdot,\cdot)$ is the beta function and $n_l$ is the number of workers in the group $S_l$.
\end{proposition}

\begin{IEEEproof}
Let $P_{e,\underline{p}}$ denote the misclassification probability given the reliabilities of the $N$ workers. Then, if $u_j$ denotes the 
bit sent by the worker $j$ and the global decision is made using the Hamming distance criterion,
\begin{equation}
P_{e,\underline{p}}=\frac{1}{M}\sum_{\underline{i},l}P(\underline{u}=\underline{i}|H_l)C_{\underline{i}}^l\mbox{.}
\end{equation}

Although local decisions are dependent, they are conditionally independent given the group assignments.
\begin{eqnarray}
&&P(\underline{u}=\underline{i}|H_l)=\sum_{\underline{s}}P(\underline{u}=\underline{i}|\underline{S}=\underline{s},H_l)P(\underline{S}=\underline{s}|H_l)\nonumber\\
&&=\sum_{\underline{s}}\left[\prod_{j=1}^N P(u_j=i_j|S_j=s_j, H_l)\right]P(\underline{S}=\underline{s}|H_l).\nonumber
\end{eqnarray}

Further,
\begin{eqnarray}
&&P(u_j=i_j|S_j=s_j, H_l)\nonumber\\
&=&i_jP(u_j=1|S_j=s_j,H_l)+(1-i_j)P(u_j=0|S_j=s_j,H_l)\nonumber\\
&=&(1-i_j)+(2i_j-1)P(u_j=1|S_j=s_j,H_l)\nonumber\\
&=&(1-i_j)+(2i_j-1)\sum_{k=1}^Ma_{kj}P(y_j=k|S_j=s_j,H_l)\nonumber\\
&=&(1-i_j)+(2i_j-1)\left(p_ja_{lj}+\frac{(1-p_j)}{(M-1)}\sum_{k\neq l}a_{kj}\right)\nonumber
\end{eqnarray}
and
\begin{eqnarray}
&&P(\underline{S}=\underline{s}|H_l)\nonumber\\
&=&E_{\underline{\lambda}}\left\{P(\underline{S}=\underline{s}|H_l,\underline{\lambda})\right\}\\
&=&E_{\underline{\lambda}}\left\{\prod_{j=1}^NP(S_j=s_j|\underline{\lambda})\right\}\\
&=&E_{\underline{\lambda}}\left\{\prod_{j=1}^N\lambda_{s_j}\right\}\\
&=&E_{\underline{\lambda}}\left\{\prod_{l=0}^{L-1}\left(\lambda_l\right)^{n_l}\right\}\\
&=&E_{\underline{\gamma}}\left\{\gamma_0^{n_0}\prod_{l=1}^{L-1}\left[\gamma_l\prod_{g=1}^{l-1}(1-\gamma_g)\right]^{n_l}\right\}\\
&=&E_{\underline{\gamma}}\left\{\gamma_0^{n_0}\prod_{l=1}^{L-1}\left[\gamma_l^{n_l}\prod_{g=1}^{l-1}(1-\gamma_g)^{n_l}\right]\right\}\\
&=&E_{\underline{\gamma}}\left\{\gamma_0^{n_0}\gamma_1^{n_1}(1-\gamma_0)^{n_1}\gamma_2^{n_2}(1-\gamma_1)^{n_2}(1-\gamma_0)^{n_2}\cdots\right\}\nonumber\\
&=&E_{\underline{\gamma}}\left\{\prod_{l=0}^{L-1}\gamma_l^{n_l}(1-\gamma_l)^{(N-\sum_{g=0}^l n_g)}\right\}
\end{eqnarray}
Since $\underline{\gamma}$ are independently drawn from the beta distribution $\gamma_l\sim \beta(1,\kappa)$,
\begin{eqnarray}
P(\underline{S}=\underline{s}|H_l)&=&\prod_{l=0}^{L-1}E\left[\gamma_l^{n_l}(1-\gamma_l)^{(N-\sum_{g=0}^l n_g)}\right]\\
&=&\prod_{l=0}^{L-1}\frac{\beta\left(n_l+1,N+\kappa-\sum_{g=0}^ln_g\right)}{\beta(1,\kappa)}\\
&=&\frac{1}{[\beta(1,\kappa)]^L}\prod_{l=0}^{L-1}\beta\left(n_l+1, N+\kappa-\sum_{g=0}^l n_g\right)
\end{eqnarray}
\end{IEEEproof}

For workers without peer-dependent reward, the reliabilities $p_j$ are independent random variables with mean $r_{s_j}$. Using the fact $E[r_{s_j}]=\mu$ yields the desired result.

Next, we introduce the peer-dependent reward scheme.

\begin{proposition}
\label{prop:error_prob_dep_corr}
Consider an $(N, M, \mu, \rho, \kappa)$ crowdsourcing system. The expected misclassification probability using code matrix $\mathbf{A}$ is:
\begin{multline}
\label{eq:err_coding_ci_pdr}
P_e(\mu,\rho, \kappa)=\frac{1}{M}\sum_{\underline{i},l,\underline{s}}C_{\underline{i}}^lP(\underline{S}=\underline{s}|H_l)\prod_{j=1}^{\frac{N}{2}}\bigg[(1-i_j)(1-i_{j_p})+(1-i_j)(2i_{j_p}-1)\left(\mu a_{lj_p}+\frac{(1-\mu)}{(M-1)}\sum_{k\neq l}a_{kj_p}\right)\\
+(1-i_{j_p})(2i_{j}-1)\left(\mu a_{lj}+\frac{(1-\mu)}{(M-1)}\sum_{k\neq l}a_{kj}\right)+(2i_j-1)(2i_{j_p}-1)\Bigg((\rho +\mu^2)a_{lj}a_{lj_{p}}+\frac{\mu-(\rho+\mu^2)}{(M-1)}\\
\left(a_{lj}\sum_{k\neq l}a_{kj_p}+a_{lj_p}\sum_{k\neq l}a_{kj}\right)+\frac{4r}{M^2}\sum_{k\neq l}a_{kj_p}\sum_{k\neq l}a_{kj}\Bigg)\bigg]
\end{multline}
where $r=\left(\frac{M}{2(M-1)}\right)^2[(1-\mu)^2+\rho]$, $\underline{i}=[i_1,\cdots,i_N] \in \{0,1\}^N$ is the received codeword, $\underline{s}=[s_1,\cdots,s_N] \in \{0,\cdots, L-1\}^N$ is the group assignment, $L$ is the number of groups and $C_{\underline{i}}^l$ is the cost associated with a global decision $H_l$ when the received vector is $\underline{i}$. This cost is given in \eqref{eq:cost}. 
The probability $P(\underline{S}=\underline{s}|H_l)$ is a function of the concentration parameter $\kappa$ and is:
\begin{equation}
P(\underline{S}=\underline{s}|H_l)=\frac{1}{[\beta(1,\kappa)]^L}\prod_{l=0}^{L-1}\beta\left(n_l+1, N+\kappa-\sum_{g=0}^l n_g\right)
\end{equation}
where $n_l$ is the number of workers in the group $S_l$.
\end{proposition}

\begin{IEEEproof}
The proof proceeds similarly to the proof of Prop.~\ref{prop:error_prob_dep_uncorr}. However, since the workers are paired, they are correlated according to \eqref{eq:cov}. This gives us
\begin{align*}
E\left[\prod_{j=1}^N P(u_j=i_j|S_j=s_j, H_l)\right]&=\prod_{j=1}^{N/2}E\left[P(u_j=i_j|S_i=s_j, H_l)P(u_{j_p}=i_{j_p}|S_{j_p}=s_{j_p}, H_l)\right] \\
&=\prod_{j=1}^{N/2}E\Bigg[\left((1-i_j)+\left(p_j a_{lj} + \frac{(1-p_j)}{(M-1)}\sum_{k\neq l}a_{kj}\right)(2i_j-1)\right)\\&\left((1-i_{j_p})+\left(p_{j_p} a_{lj_p} + \frac{(1-p_{j_p})}{(M-1)}\sum_{k\neq l}a_{kj_p}\right)(2i_{j_p}-1)\right)\Bigg]
\end{align*}
Using the above equation, correlation structure \eqref{eq:cov}, and the definition of $r=\left(\frac{M}{2(M-1)}\right)^2[(1- \mu)^2+ \rho]$ we get the desired result.
\end{IEEEproof}

Expressions for the majority approach can be derived as a special case of the coding-based approach by substituting the appropriate code matrix. We skip the details for the sake of brevity.

\subsection{Performance Evaluation}
\begin{figure}
\centering
\includegraphics[width = 3.5in]{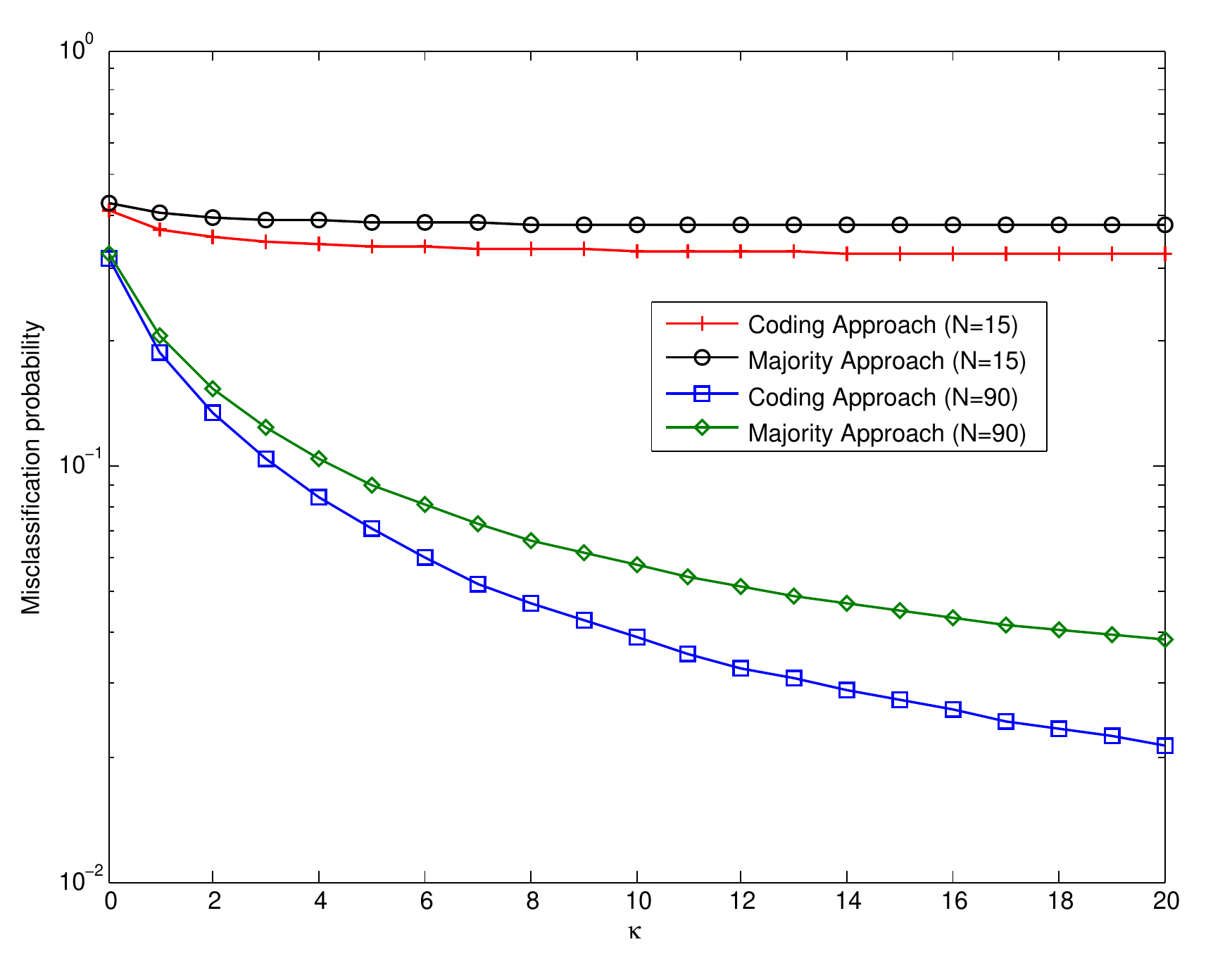}
\caption{Misclassification probability as a function of $\kappa$ using coding- and majority-based approaches with the Beta($\alpha=0.5$, $\beta=0.5$) model, $(M=8)$.}
\label{fig:err_dep_uncorr}
\end{figure}

Having analyzed the system with dependent observations among crowd workers, we evaluate the performance of the system and analyze the effect of the concentration parameter $\kappa$. Fig.~\ref{fig:err_dep_uncorr} shows the performance of the system as a function of the concentration parameter ($\kappa$). Note that a high value of $\kappa$ implies independent observations while $\kappa=0$ implies completely dependent observations (all workers have a single source of information). The plots in Fig.~\ref{fig:err_dep_uncorr} are for a system with uncorrelated workers (no peer-dependent reward scheme) performing an $8$-ary classification ($M=8$). 

As expected, we can observe that the system performance improves when the observations become independent. Also, the performance gap between majority- and coding-based approaches increases as the observations become independent. When the observations are dependent, all workers have similar observations on average, so posing similar questions (majority-based approach) will perform similarly to posing diverse questions (coding-based approach). On the other hand, when the observations are independent, it is more informative to pose diverse questions to these workers as done in the coding-based approach. Therefore, we infer that \textit{diversity is good} and the benefit of using coding increases when the observations are more diverse.

\begin{figure}
\centering
\includegraphics[width = 3.5in]{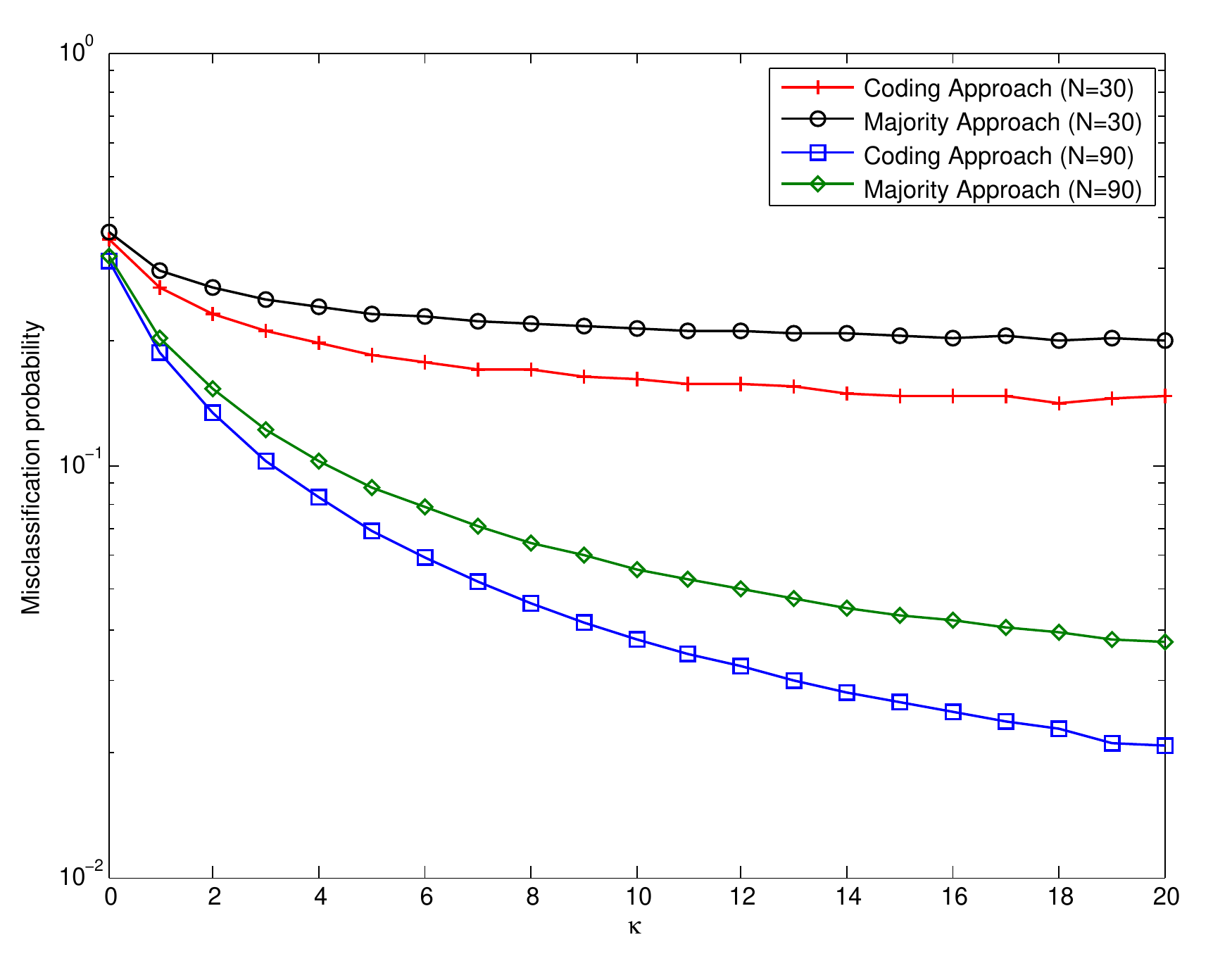}
\caption{Misclassification probability as a function of $\kappa$ using coding- and majority-based approaches with the Beta($\alpha=0.5$, $\beta=0.5$) model and $\rho=-0.5$, $(M=8)$.}
\label{fig:err_dep_corr}
\end{figure}

Similar observations can also be made for a system consisting of peer-dependent reward scheme with dependent observations. Fig.~\ref{fig:err_dep_corr} shows the performance of such a system using both majority- and coding-based approaches. The plots are for a system with $N$ crowd workers performing an $8$-ary classification task ($M=8$). The group reliability distribution is assumed to be beta distribution and the correlation parameter is $\rho_{corr}=-0.5$.

\section{Conclusion}
\label{sec:conc}
We have proposed the use of coding for reliable classification using unreliable crowd workers. Using different crowdsourcing models, we have shown that coding-based methods in crowdsourcing can more efficiently use human cognitive energy over traditional majority-based methods.  Since minimum Hamming distance decoding is equivalent to MAP decoding in this setting, the anonymity of unreliable crowd workers is not a problem. We have shown that better crowds yield better performance. In other words, crowds which have higher reliabilities on an average perform better than the crowds with lower reliabilities. Although dependent observations have been considered in typical sensor networks, human decision makers in a crowdsourcing system give rise to multiple levels of dependencies. Our work provides a mathematical framework to analyze the effect of such dependencies in crowdsourcing systems and provides some insights into the design aspects. By considering a model with peer-dependent reward scheme among crowd workers, we showed that pairing among workers can improve performance. Note that this aspect of rewards is a distinctive feature of crowdsourcing systems. We also showed that diversity in the system is desirable since the performance degrades when the worker observations are dependent. 

The benefits of coding are especially large for applications where the number of classes is large, such as fine-grained image classification for building encyclopedias like Visipedia\footnote{http://www.vision.caltech.edu/visipedia/}. In such applications, one might need to classify among more than $161$ breeds of dogs or $10000$ species of birds. Designing easy-to-answer binary questions using the proposed scheme greatly simplifies the workers' tasks. Going forward, many further questions may be addressed; some examples are as follows. Can better cognitive and attentional models of human crowd workers provide better insight and design principles?  When considering average misclassification probability, the ordering of crowd quality depends only on a first-moment characterization; what about finer characterizations of system performance? One can also design the number of paired workers in the peer-dependent reward scheme to optimize the system performance. In the current work, we designed the code matrices with no prior knowledge of the crowd parameters. In the future, we can consider a setup where the crowd parameters such as the covariance and/or dependence parameters are known to further improve the system performance. 

\bibliographystyle{IEEEtran}
\bibliography{abrv,conf_abrv,aditya_lib}

\end{document}